\documentclass[11pt]{article}
\pdfoutput=1
\usepackage{jcapmod}
\usepackage{booktabs}
\usepackage[english]{babel}
\usepackage{amsmath,amssymb,amsbsy,amstext, amsthm, simplewick}
\usepackage{hyperref}
\usepackage{graphicx}
\usepackage{amsfonts}
\usepackage{amssymb}
\usepackage{upgreek}
 \usepackage{exscale,relsize}
 \usepackage[makeroom]{cancel}
\usepackage{soul}
\usepackage[utf8]{inputenc}
\RequirePackage{color}

\usepackage{colortbl}
\definecolor{green2}{cmyk}{0, 1, 0.5, 0}
\definecolor{lightgreen}{cmyk}{0.2, 0, 0.2, 0.2}
\definecolor{lightgray}{cmyk}{0.1,0.2,0,0.1}
\definecolor{lightgray2}{cmyk}{0.4,0.4,0,0.8}
\definecolor{black}{cmyk}{1.0,1.0,1.0,1.0}

\allowdisplaybreaks[1]

\usepackage{colortbl}
\definecolor{lightgreen}{cmyk}{0.2, 0, 0.2, 0.2}
\definecolor{lightgray}{cmyk}{0.1,0.2,0,0.1}
\definecolor{lightgray2}{cmyk}{0.1,0.1,0,0.1}

\makeatletter
\newlength{\apb@width}
\newcommand{\autoparbox}[2][c]{\settowidth{\apb@width}{#2}\parbox[#1]{\apb@width}{#2}}

\makeatother

\numberwithin{equation}{section}

\def\beq{\begin{equation}}
\def\eeq{\end{equation}}

\def\bea{\begin{eqnarray}}
\def\eea{\end{eqnarray}}
\def\eg{{\it e.g.~}}

\def\ie{{\it i.e.~}}
\def\d{{\rm d}}

\def\mP{\mathcal{P}}
\def\mH{\mathcal{H}}

\def\d{{\rm d}}

\def\nn{\nonumber}

\def\sgm{\sigma}

\def\del{\partial}
\def\Mp{M_{\rm pl}}
\def\ep{\epsilon_{\phi}}
\def\es{\epsilon_{\sgm}}

\def\fr{\frac}

\def\0{{\boldsymbol 0}}

\def\fnl{f_{\mathsmaller{\rm NL}}}
\def\fr{\frac}

\DeclareSymbolFont{extraup}{U}{zavm}{m}{n}
\DeclareMathSymbol{\varheart}{\mathalpha}{extraup}{86}
\DeclareMathSymbol{\vardiamond}{\mathalpha}{extraup}{87}

\DeclareRobustCommand{\SkipTocEntry}[4]{}

\begin{document}

\begin{titlepage}

\setcounter{page}{1} \baselineskip=15.5pt \thispagestyle{empty}

\bigskip\

\vspace{1cm}
\begin{center}

{\fontsize{20}{28}\selectfont  \sffamily \bfseries On Synthetic Gravitational Waves from Multi-field Inflation}

\end{center}

\vspace{0.2cm}

\begin{center}
{\fontsize{13}{30}\selectfont Ogan Ozsoy$^{\bigstar\varheart}$}
\end{center}

\begin{center}

\vskip 8pt
\textsl{$^\bigstar$ College of Science, Swansea University, Swansea, SA2 8PP, UK}\\
\textsl{$^{\varheart}$ Department of Physics, Syracuse University, Syracuse, NY 13244, USA}
\vskip 7pt

\end{center}

\vspace{1.2cm}
\hrule \vspace{0.3cm}
\noindent {\sffamily \bfseries Abstract} \\[0.1cm]
We revisit the possibility of producing observable tensor modes through a continuous particle production process during inflation. Particularly, we focus on the multi-field realization of inflation where a spectator pseudoscalar $\sgm$ induces a significant amplification of the ${\rm U}(1)$ gauge fields through the coupling $\propto \sgm F_{\mu\nu}\tilde{F}^{\mu\nu}$. In this model, both the scalar $\sgm$ and the Abelian gauge fields are gravitationally coupled to the inflaton sector, therefore they can only affect the primordial scalar and tensor fluctuations through their mixing with gravitational fluctuations. Recent studies on this scenario show that the sourced contributions to the scalar correlators can be dangerously large to invalidate a large tensor power spectrum through the particle production mechanism. In this paper, we re-examine these recent claims by explicitly calculating the dominant contribution to the scalar power and bispectrum. Particularly, we show that once the current limits from CMB data are taken into account, it is still possible to generate a signal as large as $r \approx 10^{-3}$ and the limitations on the model building are more relaxed than what was considered before.
\vskip 10pt
\hrule

\vspace{0.6cm}
 \end{titlepage}

 \tableofcontents
 
\newpage
\section{Introduction}
In single scalar field models of inflation, it is commonly stated that a detection of primordial gravitational waves (GW's) provides us both  the energy scale at which inflation takes place and a lower bound on the field space excursion of the inflaton. These direct relationships can be understood in terms of the strength of the signal which is parametrized by the tensor-to-scalar ratio $r \equiv \Delta_t^2 /\Delta_s^2$~:
\beq\label{EI}
H_{\rm inf} \simeq 2.8 \times 10^{-5} \left(\fr{r}{0.07}\right)^{1/2} \Mp ,~~~~ \fr{\Delta\phi}{\Mp} \gtrsim 5.6\left(\fr{r}{0.07}\right)^{1/2} .
\eeq
This is why the observation of GW's from inflation is one of the main objectives of the current and upcoming CMB experiments that are expected to be sensitive to a signal at the level of, $r \lesssim 10^{-3}$ \cite{Abazajian:2016yjj, Kamionkowski:2015yta}.

The observation of GW's of primordial origin does not ensure that the expressions in \eqref{EI} is valid as it relies on the assumption that metric fluctuations originate from quantum vacuum fluctuations during inflation. In principle, it is possible to invalidate this result by considering additional fields during inflation that are not in their vacuum configuration \cite{Cook:2011hg,Senatore:2011sp}. However, recent studies have shown that it can be challenging to realize such mechanisms without spoiling the successful predictions of inflation. The main issue here is that the source of the GW's also interacts with the visible sector directly with a coupling that is stronger than the gravitational strength, thus affecting the scalar perturbations to a high degree. As the sources are non-vacuum contributions, this in general leads to large non-gaussian statistics for the scalar fluctuations especially if we insist on a large tensor power spectrum that is dominated by these sources \cite{Barnaby:2010vf, Mirbabayi:2014jqa}.

A simple way of avoiding these conclusions is to consider a hidden sector that is only gravitationally coupled to inflaton (therefore reducing the effects of the sources to a level consistent with observations). Ref. \cite{Barnaby:2012xt} considered a possible mechanism satisfying this criteria where a slowly-rolling\footnote{See also a modified version of this scenario where the scalar field experiences a transient roll, namely $\dot{\sgm}^2 \neq 0$ and $ \dot{\sgm}^2 \ll 2H^2\Mp^2 $ only for a limited time \cite{Namba:2015gja,Peloso:2016gqs}.} spectator scalar $\sgm$ amplifies Abelian gauge fields (through an axion-like coupling, \ie $\sgm F_{\mu\nu}\tilde{F}^{\mu\nu}$) which in turn act as an alternative\footnote{Other mechanisms for generating GW's alternative to the standard vacuum fluctuations include the amplification of chiral tensor modes in inflationary models with non-abelian gauge fields \cite{Dimastrogiovanni:2012ew,Adshead:2013qp,
Namba:2013kia,Obata:2014loa,Obata:2016tmo,Maleknejad:2016qjz,Adshead:2017hnc}, amplification of tensor modes by spectator fields \cite{Biagetti:2013kwa,Biagetti:2014asa,Fujita:2014oba}, modification of tensor dispersion relation \cite{Cannone:2014uqa,Cannone:2015rra} and breaking of space diffeomorphisms in the effective field theory approach to inflation \cite{Bartolo:2015qvr}.} source for GW's. It was shown there that the interaction $\delta A + \delta A \to \delta\phi$ induced by the gravitational fluctuations leads to a negligible contribution to the scalar correlators hence allowing to a visible sourced GW signal. The model in \cite{Barnaby:2012xt} was further investigated in  \cite{Ferreira:2014zia}. There it was shown that the dominant channel that feeds into the curvature correlators is due to the conversion of the spectator fluctuations $\delta\sgm$ (that is sourced by the gauge fields) to the inflaton fluctuations $\delta\phi$ through the mass mixing between $\delta\phi$ and $\delta\sgm$ induced by the gravity: namely the process $\delta A + \delta A \to \delta\sgm \to \delta \phi$ (See also the discussion in \cite{Mirbabayi:2014jqa}). The authors in \cite{Ferreira:2014zia} then concluded that in order to avoid excess power in the scalar fluctuations, the spectator field $\sgm$ should decay\footnote{The mass mixing between the scalar fluctuations are proportional to $\propto \sqrt{\es}$ and therefore it is absent when the spectator decays, \ie $\es \to 0$.} in less than two e-folds $N_{\sgm} \lesssim 2$ in order to grant for observably large tensors.

In light of these issues mentioned in the literature, in this work, we re-investigate the multi-field model of \cite{Barnaby:2012xt}. For this purpose, focusing on spatially flat gauge, we explicitly calculate the sourced scalar correlators in the model and show that a factor of $(\ep N_{\sgm})^2$ enters in the scalar power spectrum, and a factor of $(\ep N_{\sgm})^3$ in the bispectrum. Although these findings are in agreement with \cite{Ferreira:2014zia} qualitatively, we found that the model can still account for observably large tensor modes ($r \approx 10^{-3}-10^{-4}$) at CMB scales\footnote{Apart from the observational consequences at CMB scales, this scenario have a rich variety of observational implications such as observable GW's at interferometer scales \cite{Cook:2011hg,Barnaby:2011qe,Domcke:2016bkh,Garcia-Bellido:2016dkw}, parity violation at CMB \cite{Cook:2013xea,Sorbo:2011rz,Shiraishi:2013kxa} and at interferometer scales \cite{Crowder:2012ik} and primordial black holes \cite{Linde:2012bt,Erfani:2015rqv} in different regions of its parameter space.} once the Planck limits on non-gaussianities are respected while the spectator is allowed to roll for $N_\sgm = 7-10$ before it decays.

The remainder of the paper is as follows. In Section \ref{Sec2}, we give a brief overview of the model including the background evolution and the particle production in the gauge field sector. In Section \ref{Sec3}, we study the dynamics of scalar and tensor fluctuations sourced by the gauge field and identify the important observables at CMB scales. In Section \ref{Sec4}, we summarize our results for scalar and tensor correlators and discuss in detail the phenomenology that might arise in light of the constraints on non-gaussianity at CMB scales and from model building. In Section \ref{Sec5}, we present our conclusions. This works is supplemented by three appendices. In Appendix A, we present the all the interaction terms in the model using the ADM formalism and investigate in detail all the source terms in the equations of motion of scalar and tensor fluctuations. In Appendix B, we review the details of gauge field production. In Appendix C, we present the calculational details for the scalar power spectrum and bispectrum. 
\\

{\bf Notation and conventions.} We will use natural units, $\hbar=c=1$, with reduced Planck mass $\Mp^2 =(8\pi G)^{-1}$. Our metric signature is mostly plus sign $(-,+,+,+)$. Greek indices stand for space-time coordinates, while Latin indices denote spatial coordinates. Overdots and primes on time dependent quantities will denote derivatives with respect to coordinate time $t$ and conformal time $\tau$, respectively. During inflation, we take $a(\tau)= 1/(-H\tau)^{1+\epsilon}$ with $H$ is the physical Hubble rate, while the comoving one is denoted by $\mathcal{H} = aH$.   
\section{The Model}\label{Sec2}
We consider the model described by the following matter Lagrangian \cite{Barnaby:2012xt},
\beq\label{Lm}
\mathcal{L}_m =  -\underbrace{\fr{1}{2}(\del\phi)^2 - V_{\phi}(\phi)}_\text{Inflaton Sector}-\underbrace{\fr{1}{2}(\del\sigma)^2 - V_{\sgm}(\sgm)-\fr{1}{4}F_{\mu\nu}F^{\mu\nu}-\fr{\sgm}{4f}F_{\mu\nu}\tilde{F}^{\mu\nu},}_\text{Hidden Sector}
\eeq
where $\phi$ is the inflaton and the hidden sector includes the pseudoscalar $\sigma$, the gauge field $A_\mu$ and their interaction through the Chern-Simons term. Here, $V_\phi(\phi)$ and $V_\sgm(\sgm)$ are the potential of the inflaton and the pseudoscalar, whereas the gauge field strength tensor and its dual are defined by $F_{\mu\nu} = \del_\mu A_\nu - \del_\nu A_\mu$ and $\tilde{F}^{\mu\nu}\equiv \eta^{\mu\nu\rho\sigma} F_{\rho\sigma}/(2\sqrt{-g})$ where alternating symbol $\eta^{\mu\nu\rho\sigma}$ is $1$ for even permutation of its indices, $-1$ for odd permutations, and zero otherwise.
\subsection{Background dynamics}\label{sec21}
We assume that $\phi$ and $\sgm$ take homogeneous vacuum expectation  values (vev), $\phi_{_0}(t)$ and $\sgm_{_0}(t)$, respectively and the background spacetime takes the flat FLRW form: ${\rm diag}(-1,a^2,a^2,a^2)$. Then the equations of motion for the background fields can be obtained from \eqref{Lm} 
\bea
\ddot{\phi}_{_0} + 3H\dot{\phi}_0 + V_{\phi}'(\phi_{_0}) &=&0,\\
\label{beom2}\ddot{\sgm}_{_0} + 3H\dot{\sgm}_0 + V_{\sgm}'(\sgm_{_0}) &=&\fr{1}{f}\langle \vec{E}.\vec{B} \rangle, 
\eea 
where $H$ is the physical Hubble rate. Here, we assumed that the gauge fields have no vev and hence we introduced their effects on the background evolution as higher order expectation values using the mean field approximation. These equations can be combined with Friedmann equations to determine the full background evolution 
\bea
\nn 3H^2 \Mp^2 &=& \frac{1}{2}  \dot{\phi}_{_0}^2  +\frac{1}{2} \dot{\sgm}_{_0}^2 + V_{\phi}(\phi_{_0}) +V_{\sgm}(\sgm_{_0}) + \fr{1}{2}\langle \vec{E}^2 +\vec{B}^2 \rangle,\\
-2\dot{H} \Mp^2 &= &  \dot{\phi}_{_0}^2 + \dot{\sgm}_{_0}^2, 
\eea
As in \cite{Barnaby:2012xt}, we consider a multi-field inflationary setup where both background fields are rolling slowly at approximately constant velocity\footnote{Up to corrections of the order of the change in slow-roll parameters of the fields which we assume to be small $\dot{\epsilon}/\epsilon H \ll 1$}, \ie $|\dot{\phi}_{_0}| \ll \sqrt{2}H\Mp$, $|\dot{\sgm}_{_0}| \ll \sqrt{2}H\Mp$ where the background dynamics are mainly controlled by the inflaton, $\rho_\sgm \ll 3H^2\Mp^2 \approx V(\phi)$. Note that these conditions do not necessarily imply a hierarchy between the velocities (or for the ratio of the slow-roll parameters) of the scalar fields. In addition to the assumptions above, we will require that $\sgm$ rolls by a small amount of e-folds $N_\sgm$ compared to the total duration of inflation\footnote{In Section \ref{Sec3} and \ref{Sec4}, we will show that this requirement on the background model essentially originates in order to reconcile with cosmological observations.} and settles to the minimum of its potential long before inflation ends. This can be easily realized by choosing suitable initial conditions together with a potential $V_\sgm$ that has a larger slope compared to the slow-roll potential of the inflaton $V_\phi$.  

In order to avoid back-reaction of the produced gauge quanta on the expansion of the universe, we require the source of the particle production to be larger then the energy density contained in the gauge fields. Using the expression in \eqref{EV}, this gives
\beq\label{mbc}
\boxed{\dot{\sgm}_{_0}^2 > \langle \vec{E}^2  +\vec{B}^2 \rangle~~~ \to~~~ 
\fr{H^2}{\dot{\sgm}_{_0}} < 60 \xi^{3/2} e^{-\pi\xi}.}
\eeq
On the other hand, the gauge field should have a negligible contribution on the background evolution of $\sgm$, which requires $|U'_\sgm|<f^{-1}|\langle\vec{E}.\vec{B}\rangle$ in \eqref{beom2}. Using the first expression in \eqref{EV}, this condition can be shown to be sub-dominant compared to the one we consider in equation \eqref{mbc}. 

In this paper, we will work in the regime of negligible back-reaction. In this case, the background model we described above can be approximated by the following equations,
\bea
\ddot{\phi}_{_0} + 3H\dot{\phi}_0 + V_{\phi}'(\phi_{_0}) &=&0,\\
\label{beom2}\ddot{\sgm}_{_0} + 3H\dot{\sgm}_0 + V_{\sgm}'(\sgm_{_0}) &\simeq&0
\eea
and
\bea
\nn 3H^2 \Mp^2 &\approx & V_{\phi}(\phi_{_0}),\\
-2\dot{H} \Mp^2 &= &  \dot{\phi}_{_0}^2 + \dot{\sgm}_{_0}^2. 
\eea

\subsection{Gauge Field Production}
In this section, we briefly review the gauge field production\footnote{Some details on particle production can be found in the Appendix B.} in the background model described above. The equation of motion for the gauge field can be obtained by varying the action in \eqref{Lm} in Coulomb gauge (see Appendix A),  
\beq\label{EMG}
A_i'' - \vec{\nabla}^2 A_i - \fr{ a \dot{\sgm}_{_0}}{f} ~\epsilon_{ijk}~ \del_j A_k = 0.
\eeq
We decompose the gauge field $A_i$ in terms of the annihilation and creation operators in the usual way,
\beq\label{DGF}
A_i(\tau, \vec{x}) = \sum_{\lambda = \pm} \int \fr{\d^3 k}{(2\pi)^{3/2}}\left[\epsilon^{\lambda}_i(\vec{k})
A_\lambda(\tau,\vec{k})\hat{a}_\lambda(\vec{k})e^{i\vec{k}.\vec{x}} + h.c.\right],   
\eeq
where the helicity vectors obey $k_i \epsilon^{\pm}_i = 0$, $\epsilon_{ijk}~ k_j ~\epsilon^{\pm}_k = \mp i k \epsilon^{\pm}_i$, $\epsilon^{\pm}_i\epsilon^{\pm}_i = 0$ and $\epsilon^{\pm}_i \epsilon^{\mp}_i =1$.

The annihilation/creation operators satisfy
\beq\label{cr}
\left[\hat{a}_\lambda(\vec{k}),\hat{a}^\dagger_{\lambda'}(\vec{k}')\right] = \delta_{\lambda\lambda'} \delta^{(3)}(\vec{k}-\vec{k}').
\eeq
Plugging the decomposition in \eqref{DGF} into \eqref{EMG}, we write the equation of motion for the mode functions of the gauge field as 
 
\beq\label{MEA}
{A}_\pm'' + \left(k^2 \pm \fr{2k\xi}{\tau}\right)A_\pm =0, 
\eeq
where prime denotes a derivative with respect to conformal time $\d\tau =\d t/a $ and we have defined a dimensionless measure of field velocity $\xi \equiv \dot{\sgm}_{_0}/2Hf$. From equation \eqref{MEA}, we see that positive helicity modes of gauge field $A_+$ exhibit tachyonic instability for modes satisfying $-k\tau < 2\xi$ where we assumed $\dot{\sgm}_{_0} >0$ without loss of generality. In the case that $\xi$ varies slowly, \ie $\dot{\xi}/\xi H \ll 1$, the growth of the fluctuations can be shown to satisfy
\beq\label{MEAS}
A_+  \simeq \fr{1}{\sqrt{2k}} \left(\fr{-k\tau}{2\xi}\right)^{1/4} e^{\pi\xi -2\sqrt{-2\xi k\tau}}
\eeq 
for modes satisfying $(8\xi)^{-1}\lesssim -k\tau \lesssim 2\xi$ \cite{Barnaby:2010vf}. These modes account for the most of the power contained in the gauge field fluctuations. Note that for $\xi > 1/4$ this phase space is non-vanishing. In this work, we will work in the regime where $\xi \gtrsim \mathcal{O}(1)$ to account for efficient particle production. Since only the positive helicity modes are amplified, we will use the following expression for the gauge fields in the rest of this paper
\beq
A_i(\tau, \vec{x}) =\int \fr{\d^3 k}{(2\pi)^{3/2}} e^{i\vec{k}.\vec{x}} \epsilon^{+}_i(\vec{k}) A_+ (k,\tau)\left[\hat{a}_+(\vec{k})+\hat{a}^{\dagger}_+(-\vec{k})\right],
\eeq
where $A_+$ is given in \eqref{MEAS}. For future reference, we also define ``Electric'' and ``Magnetic" fields\footnote{Although $A_i$ is not neccessarily the Standard Model $U(1)$ gauge field, we will continue to use the standard electromagnatic notation for convenience.} which are related to the auxiliary potential $A_i$ as
\beq\label{EMF}
E_i = -\fr{1}{a^2}~ A_i' , ~~~ B_i = \fr{1}{a^2}~\epsilon_{ijk}~\del_j A_k, 
\eeq 
and thus we have
\bea\label{EaB}
\nn E_i(\tau, \vec{x}) &=& -\fr{1}{a^2(\tau)} \int\fr{\d^3 k}{(2\pi)^{3/2}} e^{i\vec{k}.\vec{x}} \epsilon^{+}_i(\vec{k}) A'_+ (k,\tau)\left[\hat{a}_+(\vec{k})+\hat{a}^{\dagger}_+(-\vec{k})\right],\\
B_i(\tau, \vec{x}) &=& \fr{1}{a^2(\tau)} \int\fr{\d^3 k}{(2\pi)^{3/2}} e^{i\vec{k}.\vec{x}} \epsilon^{+}_i(\vec{k})~ |\vec{k}|~ A_+ (k,\tau)\left[\hat{a}_+(\vec{k})+\hat{a}^{\dagger}_+(-\vec{k})\right].
\eea 
Using the expressions in \eqref{EaB}, some important expectation values including electric and magnetic fields can be calculated analytically (see Appendix B):
\beq\label{EV}
\langle\vec{E}.\vec{B}\rangle\simeq -2.4\times 10^{-4}\fr{H^4}{\xi^4} e^{2\pi\xi}, ~~ \fr{1}{2}\langle\vec{E}^2+\vec{B^2}\rangle \simeq 1.4 \times 10^{-4} \fr{H^4}{\xi^3} e^{2\pi\xi} . 
\eeq
These expectation values are useful in identifying back-reaction effects of the produced particles on the background evolution as well as in constraining the allowed parameter space in the model as we will turn later on. 

\section{Dynamics of the Fluctuations}\label{Sec3}
The gravitational coupling between the inflaton and the hidden sector fields ($\sgm$ and $A_i$) induces source terms in the equation of motion of the inflaton fluctuations. We therefore expect to have additional contributions to the curvature perturbation besides those due to vacuum fluctuations. Moreover, the gauge field also give rise to a source term in the equation of motion of tensor modes in addition to those generated by quantum vacuum fluctuations of the metric. In this section, we will therefore focus on the correlators of scalar curvature perturbations and those involving tensor metric perturbations in the presence of gauge fields. Some of the details regarding the dynamics of the scalar and tensor fluctuations can be found in Appendix A.  
\subsection{Scalars}
We start our discussion with scalar fluctuations as their dynamics are much more involved compared to the tensors. In flat gauge, switching to conformal time $\d\tau = \d t/a$, the equation of motion of the inflaton fluctuation $\delta \phi$ in momentum space reads 
\beq\label{eom1}
\delta \phi'' + 2\mH\delta\phi' + (k^2 + a^2 V''_\phi)~\delta\phi-\left(\fr{a^2 \phi^{'2}_{_0}}{\mH}\right)\fr{\delta\phi}{a^2\Mp^2}-\left(\fr{a^2 \phi^{'}_{_0}\sgm'_{_0}}{\mH}\right)\fr{\delta\sgm}{a^2\Mp^2}\simeq J^{(\phi)}_{\rm GI}(\vec{k},\tau),
\eeq
where the gravitationally induced source term is given by a convolution in momentum space,
\beq\label{seom1}
J^{(\phi)}_{\rm GI}(\vec{k},\tau)=-\fr{\phi_{_0}'}{2\mathcal{H}\Mp^2}\fr{a^2}{k^2}~k_ik_j\int \fr{\d^3 p}{(2\pi)^{3/2}}E_i(\vec{k}-\vec{p},\tau)E_j(\vec{p},\tau).
\eeq
The approximate equality in \eqref{eom1} is due to the fact that ``magnetic'' fields can be neglected compared to the ``electric'' fields in the model under consideration (See the discussion in Section \ref{ESeom}).  
In \eqref{eom1}, it is important to note that metric fluctuations gave rise to a term proportional to $\delta \sgm$, whose presence is crucial for the calculation of sourced scalar correlators in this model. This is especially clear when we have a look at the e.o.m of spectator field fluctuations $\delta \sgm$:
\beq\label{eom2}
\delta \sgm'' + 2\mH\delta\sgm' + (k^2 + a^2 V''_\sgm)~\delta\sgm-\left(\fr{a^2 \sgm^{'2}_{_0}}{\mH}\right)\fr{\delta\sgm}{a^2\Mp^2}-\left(\fr{a^2 \phi^{'}_{_0}\sgm'_{_0}}{\mH}\right)\fr{\delta\phi}{a^2\Mp^2}\simeq J^{(\sgm)}(\vec{k},\tau),
\eeq
where the sourced $\delta \sgm$ in \eqref{eom2} can leak into $\delta \phi$ (via $\delta A +\delta A\to \delta \sgm\to \delta \phi$)  through the mixing term in \eqref{eom1} as far as the spectator field rolls, \ie $\sgm'_{_0}\neq 0$. This is the process  considered as dangerous in \cite{Ferreira:2014zia} especially if we require the sourced tensor power spectrum (See Section \ref{T}) to dominate over the vacuum fluctuations. Below, we will re-examine this claim in detail (see Section \ref{Sec4}). First, we focus on the coupled system of equations in \eqref{eom1} and \eqref{eom2} at leading order in slow-roll expansion. The source term in the e.o.m of $\delta \sgm$ consist of two parts and it is convinient to split it into one that arise from the direct coupling of scalar $\sgm$ to gauge fields plus a term induced by gravity:
\beq
J^{(\sgm)}(\vec{k},\tau) = J^{(\sgm)}_{\rm D}(\vec{k},\tau) + J^{(\sgm)}_{\rm GI}(\vec{k}, \tau)
\eeq
with
\bea
\label{ds}J^{(\sgm)}_{\rm D}(\vec{k},\tau) &=& \fr{a^2}{f}\int \fr{\d^3 p}{(2\pi)^{3/2}}E_i(\vec{k}-\vec{p},\tau)B_i(\vec{p},\tau),\\
J^{(\sgm)}_{\rm GI}(\vec{k},\tau) &\simeq& -\fr{\sgm_{_0}'}{2\mathcal{H}\Mp^2}\fr{a^2}{k^2}~k_ik_j\int \fr{\d^3 p}{(2\pi)^{3/2}}E_i(\vec{k}-\vec{p})E_j(\vec{p}).
\eea
Notice that for both scalar fluctuations there is a democratic source term induced by gravitational fluctuations (see \eg the equation \eqref{gi1} and \eqref{gi2} and the following discussion on source terms in Appendix A). We define the canonical fields, $u_a = a \left(\delta \phi, \delta \sgm\right)^T$ to re-write the system of equations \eqref{eom1}-\eqref{eom2} as
\beq
\left(\partial_\tau^2 +k^2 -\fr{2}{\tau^2}\right)u_a-\fr{{M}_{ab}}{\tau^2}  u_b = a(\tau)\left(
\begin{array}{c}
 J^{(\phi)}_{\rm GI} \\
 J^{(\sgm)}\\
\end{array}
\right),
\eeq  
where mass mixing between $u_\phi$ and $u_\sgm$ is given by the following matrix \cite{Byrnes:2006fr,Namba:2015gja}
\beq
M_{ab}=
  \left( {\begin{array}{cc}
   9\epsilon_\phi+3\epsilon_\sgm-3\eta_\phi & ~~~6\sqrt{\ep\es} \\ \\
   6\sqrt{\ep\es} & ~~~9\epsilon_\sgm+3\epsilon_\phi-3\eta_\phi \\
  \end{array} } \right),
\eeq
and where we have defined the slow-roll parameters
\beq
\epsilon_a \equiv \fr{\dot{\phi}_{_{0a}}^2}{2H^2\Mp^2},~~~~ \eta_a\equiv \Mp^2 \fr{V_{,aa}}{V},
\eeq
with $a=\{\phi,\sgm\}$ and $V = V_\phi + V_\sgm$. The matrix $M_{ab}$ can be diagonalized by $U M U^T = \Lambda ={\rm diag}(\lambda_\phi,\lambda_\sgm)$ where $U(\theta)$ is a rotation matrix:
\beq
U(\theta) = \left( {\begin{array}{cc}
   \cos\theta & ~~\sin\theta \\
   -\sin\theta & ~~\cos\theta \\
  \end{array} } \right). 
\eeq
To leading order in slow-roll expansion the eigenvalues of the matrix $\Lambda$ are given by
\beq
\lambda_a = \fr{3}{2}\Big[4(\epsilon_\phi+\epsilon_\sgm)-(\eta_\phi+\eta_\sgm)\pm\sqrt{[2(\epsilon_\phi-\epsilon_\sgm)+(\eta_\sgm-\eta_\phi)]^2+ 16\epsilon_\phi\epsilon_\sgm}\Big].
\eeq
Defining the eigenvectors of the $\Lambda$ matrix by $v_a = (U^T)_{ab}u_b$, the diagonalized system can be expressed as
\beq\label{eomf0}
\Big[\partial_\tau^2 + k^2 -\fr{1}{\tau^2}\left(\mu_a^2-\fr{1}{4}\right)\Big]v_a = a(\tau)~ S_a (k,\tau)
\eeq
where $\mu_a = 3/2 + \lambda_a/3$ and we defined
\beq
S_a \equiv \left(
\begin{array}{c}
 \cos\theta~ J^{(\phi)}_{\rm GI} + \sin\theta~ J^{(\sgm)}\\
 -\sin\theta~J^{(\phi)}_{\rm GI}+ \cos \theta~J^{(\sgm)}\\
\end{array}
\right).
\eeq
As we are interested in the behavior of scalar fluctuations in the presence of gauge field sources, we wrote the particular solution for the eigenvectors in 
\eqref{eomf0}, which is collectively given by
\beq
v_a(\vec{k},\tau) = \int \d \tau'~ G^{\mu_a}_k(\tau,\tau')~ a(\tau') ~S_a(\vec{k},\tau'),
\eeq
where $G_k^{\mu_a}$ are the retarded Green functions for the homogeneous part of the equation \eqref{eomf0} labeled by the index $\mu_a$:
\beq
G^{\mu_a}_k (\tau,\tau') = \Theta(\tau-\tau')\fr{\pi}{2}\sqrt{\tau\tau'}~\Big[J_{\mu_a}(-k\tau)Y_{\mu_a}(-k\tau')-Y_{\mu_a}(-k\tau)J_{\mu_a}(-k\tau')\Big].
\eeq 
Here, $\Theta$ is the Heaviside step function and $J$ and $Y$ denote Bessel function of real argument. Ultimately, we are interested in the original variable $u_\phi = a \delta\phi$ as it is the one directly related to the curvature perturbation (See the discussion in Section \ref{DOZ}). Using the definition, $u_a = U_{ab} v_b $, the original mode function $u_\phi$ is given by
\begin{align}\label{solf}
\nn u_\phi &= \int d\tau' a(\tau')~\Big[G^{\mu_\phi}_k(\tau,\tau')\left(\cos^2\theta~ J^{(\phi)}_{\rm GI}+\cos\theta\sin\theta~ J^{(\sgm)}\right)\\
&\quad\quad\quad\quad\quad\quad\quad+ G^{\mu_\sgm}_k(\tau,\tau')\left(\sin^2\theta~ J^{(\phi)}_{\rm GI}-\cos\theta\sin\theta~ J^{(\sgm)}\right)\Big].
\end{align}
Several simplifications can be made for the calculation of $u_\phi$ in \eqref{solf}. First, since we are interested in how $\delta \sgm$ feeds into the inflaton perturbations while they are still outside the horizon, the relevant solution for $u_\phi$ is concerned with modes outside the horizon, therefore we can take $-k\tau \ll 1$ inside the Green's functions. On the other hand, source terms $J^{(\phi)}(\vec{k},\tau')$ and $J^{(\sgm)}(\vec{k},\tau')$ in \eqref{solf} are exponentially suppressed for $-k\tau'\gtrsim 1$ (See for example \eqref{mfs}) which also allows us to assume $-k\tau' \ll 1$ in the propagator $G_k^{\mu_a}$:
\beq\label{prop1}
G^{\mu_a}_k(\tau,\tau') \simeq \fr{\sqrt{\tau\tau'}}{2\mu_a} \left(\fr{\tau'}{\tau}\right)^{\mu_a}.
\eeq
Furthermore, we can use the expression for the index of the Bessel functions $\mu_a = 3/2 + \lambda_a(\epsilon,\eta)/3$ to employ a series expansion for $G^{\mu_a}_k$:
\bea
\nn G_k^{\mu_a} &\equiv& \fr{\sqrt{\tau\tau'}}{3}\left(\fr{\tau'}{\tau}\right)^{3/2}\left(1+ \fr{\lambda_a}{3}\log \fr{\tau'}{\tau}-\fr{2\lambda_a}{9}\right)\\
&\simeq&  G_k^{(3/2)}(\tau,\tau') \left(1+\fr{\lambda_a}{3}\log \fr{\tau'}{\tau}-\fr{2\lambda_a}{9}\right). 
\eea
where $G_k^{(3/2)}$ corresponds to the propagator when $\mu_a = 3/2$ in \eqref{prop1}. Notice that first two terms in the expression above give the dominant contribution, therefore we will drop the last term when calculating $u_\phi$.  Application of the arguments above to \eqref{solf} gives 
\begin{align}
\nn u_\phi(\vec{k},\tau) &\simeq \cos\theta\sin\theta~ \fr{(\lambda_\phi-\lambda_\sgm)}{3} \int d\tau' a(\tau')~ G_k^{(3/2)}  (\tau,\tau') \log \left( \fr{\tau'}{\tau}\right) J^{(\sgm)}(\vec{k},\tau')\\
\nn &~~~~+ \fr{1}{3}\left(\lambda_\phi \cos^2\theta + \lambda_\sgm \sin^2\theta \right)  \int d\tau' a(\tau')~ G_k^{(3/2)}  (\tau,\tau') \log \left( \fr{\tau'}{\tau}\right) J^{(\phi)}_{\rm GI}(\vec{k},\tau')\\
&~~~~+ \int d\tau' a(\tau')~ G_k^{(3/2)}  (\tau,\tau') J^{(\phi)}_{\rm GI}(\vec{k},\tau').
\end{align}
A final simplification can be made by noticing that the source terms $J^{(\phi)}$ and $J^{(\sgm)}$ have peaks around a small region including $\tau' \simeq -1/k$ and therefore the integrals in $d\tau'$ will be dominated by these values of $\tau'$. This allows us to approximate $\log \tau'/\tau$ appearing inside the integrals with a constant factor of $N_{\sgm}$, the number of e-folds the $\sgm$ field is rolling from the time the mode $k$ has left the horizon. With this replacement, the solutions described above corresponds to solving the following equation of motion for $u_\phi = a \delta \phi$,
\begin{align}\label{eom3}
\nn \left(\partial_\tau^2 +k^2 -\fr{2}{\tau^2}\right)u_\phi &\simeq \fr{N_{\sgm}}{3}a(\tau)\left[ \cos\theta\sin\theta~ (\lambda_\phi-\lambda_\sgm) J^{(\sgm)} + \left(\lambda_\phi \cos^2\theta + \lambda_\sgm \sin^2\theta \right) J^{(\phi)}_{\rm GI}\right]+a(\tau) J^{(\phi)}_{\rm GI}.\\ 
\end{align}
This equation includes an effective source as a combination of those for the fluctuations of the inflaton and the spectator field. We will now turn to investigate these sources in some detail.
\subsubsection{Estimation of sources in the equation of motion of $u_\phi$}\label{ESeom}
In order to compare the relative importance of the source terms in \eqref{eom3}, we first estimate the relative amplitudes of the $\vec{E}$ and $\vec{B}$, following a similar analysis in \cite{Caprini:2017vnn}. Going back to the solution in \eqref{MEAS} and using the definitions in \eqref{EMF}, we obtain the ratio of the amplitudes of electric and magnetic field as,
\beq
\fr{|\vec{E}|}{|\vec{B}|}\sim \fr{A'_+(k,\tau)}{k A(k,\tau)}\sim \sqrt{\fr{\xi}{-k\tau}} \sim \xi,
\eeq
where we have used $-k \tau \sim \xi^{-1}$ for the optimal estimate on the ratio since for modes satisfying $-\xi k\tau \gg 1$, the power of the mode functions $A_+$ are suppressed further (See \ie equation \eqref{MEAS}). 

Using the exact relation, $
\sin \theta \cos \theta (\lambda_\phi-\lambda_\sgm) = 6 \sqrt{\epsilon_\phi \epsilon_\sgm}$ and the fact that $\lambda_\phi, \lambda_\sgm \approx \epsilon_\phi$ together with $\xi \equiv \sqrt{\epsilon_\sgm/2}~(\Mp/f)$, the following estimates can be made for the amplitude of the source terms appearing in \eqref{eom3},
\bea
\label{s1} \fr{N_{\sgm}}{3}a(\tau) \cos\theta\sin\theta~ (\lambda_\phi-\lambda_\sgm) J^{(\sgm)} &\sim& a^3 N_{\sgm} \sqrt{\ep\es} \left[\fr{1}{f}[E*B]-\fr{\es}{f}[E*B]\right],\\
\label{s2} \fr{N_{\sgm}}{3}a(\tau) \left(\lambda_\phi \cos^2\theta + \lambda_\sgm \sin^2\theta \right) J^{(\phi)}_{\rm GI} &\sim& -a^3 N_{\sgm} \sqrt{\ep\es} \left[\fr{\ep}{f}[E*B]\right],\\
\label{s3} a(\tau) J^{(\phi)}_{\rm GI} &\sim& -a^3 \sqrt{\ep\es} \left[\fr{1}{f}[E*B]\right].
\eea
In the estimates above we have suppressed the tensorial structure of the sources and relate $[E*E]\sim \xi~ [E*B]$ where $[E*B]$ stands for convolutions of the form $\int \d^3q E(\vec{k}-\vec{q})B(\vec{q})$. It is clear from the expressions above that the second term inside the square brackets in \eqref{s1}, as well as the term in \eqref{s2} are higher order in slow-roll and can be neglected compared to the first term in \eqref{s1}: $N_{\sgm} \sqrt{\epsilon_\phi \epsilon_\sgm} \left[\fr{1}{f}[E*B]\right]$. Moreover, the last source term in \eqref{s3} is of the same order compared to this term for $N_{\sgm} \sim \mathcal{O}(1)$ but for $N_{\sgm} > 1$, we expect it to be sub-dominant\footnote{In fact, the effect of this term on the curvature perturbation has been already studied in detail elsewhere \cite{Barnaby:2012xt} and found to be negligible compared to the vacuum scalar fluctuations in the parameter space where the sourced tensor spectrum (See section \ref{T}) is larger compared to the vacuum contribution.}. Therefore, in the $N_{\sgm} >1$ regime, we will focus on the following equation of motion,
\beq\label{eomf}
\left(\partial_\tau^2 +k^2 -\fr{2}{\tau^2}\right)u_\phi \simeq \fr{a(\tau)^3}{f} 2 N_{\sgm} \sqrt{\ep\es} \int \fr{\d^3 p}{(2\pi)^{3/2}}E_i(\vec{k}-\vec{p},\tau)B_i(\vec{p},\tau).
\eeq  
In summary, from the perspective of the original equations of motion \eqref{eom1} and \eqref{eom2}, the solution to the equation for $u_\phi = a \delta \phi$ in \eqref{eomf} corresponds to the process where the $\delta \sgm$ fluctuations are produced through gauge fields by the dominant source term $J_{\rm D}(\vec{k},\tau)$ in \eqref{ds} and its subsequent sourcing of $\delta \phi$ through the mixing terms in \eqref{eom1}, namely the process $\delta A + \delta A \to \delta \sgm \to \delta \phi$. The estimates of various source terms above clearly shows that this is the dominant process as far as the background field $\sgm_{_0}$ continues to roll after the observable scales crosses the horizon, \ie $N_{\sgm} >1$.
\subsubsection{The curvature perturbation $\zeta$}\label{DOZ}
In the case of single field inflation, it is enough to solve equation \eqref{eomf} to evaluate the curvature perturbation through the relation $\zeta = -(\mathcal{H}/\phi_{_0}'a)u_\phi$. However, in the presence of a second scalar field $\sgm$, this relation in general does not hold. In this subsection, we will show under which conditions we can trust the simple relation above. In flat gauge, the curvature perturbation can be related to the energy density perturbation as \cite{Bassett:2005xm},
\beq\label{mz}
\zeta = -\fr{\mathcal{H}}{\rho_{0}'} \delta\rho.
\eeq
Without making any assumptions on the relative size of the kinetic terms of the scalar fields, the derivative of the background energy density is given by $\rho_{_0}' = -3\mathcal{H} (\phi_{_0}'^2 + \sgm_{_0}'^2)/a^2$. On the other hand, energy density perturbation in \eqref{mz} can be written in terms of the scalar field fluctuations using the linearized Einstein equations in flat gauge. At leading order in scalar field fluctuations, this gives \cite{Baumann:2009ds,Malik:2008im}
\beq\label{drho}
\delta\rho \simeq -\fr{3\mathcal{H}}{a^2}\left(\phi_{_0}'\fr{u_\phi}{a} + \sgm_{_0}' \fr{u_{\sgm}}{a} \right). 
\eeq    
Combining \eqref{drho} and \eqref{mz}, we obtain the expression of $\zeta$, in terms of the field perturbations
\beq\label{zz}
\zeta = -\fr{1}{1+\Theta^2}~\fr{\mathcal{H}}{\phi_{_0}' a} \left(u_\phi + \Theta~ u_\sgm\right),
\eeq
where $\Theta \equiv \sgm'_{_0}/\phi'_{_0} $. In this work, we assume that $\sgm$ rolls for a certain amount of e-folds $N_\sgm$ before it reaches to the minimum of its potential. In this case, the  energy density and pressure in the hidden sector rapidly drops to zero. In \eqref{zz}, this implies that $\Theta \to 0$ and $\delta\rho_{\sgm} \propto u_\sgm \to 0$ at the end of inflation and therefore the contribution of the hidden sector to the late time observable curvature perturbation will be negligible\footnote{During the $N_\sgm$ e-fold of slow-roll evolution, one might worry about the production of large isocurvature fluctuations along the light $\sgm$ direction in field trajectory. However, the energy density in the $\sgm$ field will decay away soon after, erasing any observable affect of these fluctuations. Similarly, we do not expect any scale dependent feature in the $\zeta$ correlators due to trapping of the field to its minimum as at this point energy density in the hidden sector will be completely negligible. We thank to an anonymous referee for a discussion on this point.} \cite{Mukohyama:2014gba}: 
\beq\label{sz}
\zeta \simeq -\fr{\mathcal{H}}{\phi_{_0}'a} u_\phi.
\eeq
In such a situation, the only impact of hidden sector fluctuations on observed curvature perturbation is through the linear mixing term they share with the inflaton perturbations (See equation \eqref{eom1}). As we discussed earlier in Section \ref{ESeom}, the main utility of this linear interaction is to convert $\delta\sgm$ sourced by the inverse decay process of gauge fields into $\delta\phi$, \ie through the process $\delta A +\delta A \to \delta\sgm \to \delta \phi$. In particular, in the presence of scalar mixing, we have showed that the inverse decay of gauge fields can affect the dynamics of $u_\phi$ through the source term in \eqref{eomf} which is effective as long as $\sgm$ rolls, \ie $\sgm_{_0}' \neq 0$ with $N_{\sgm}\neq 0$. Therefore the indirect influence of the spectator fluctuation on late time observable $\zeta$ can be effectively captured by using the standard relation in \eqref{sz} where $u_\phi$ is the solution to the mode equation given in \eqref{eomf} \cite{Namba:2015gja}.     

\subsubsection{The sourced power spectrum and bispectrum}
In order to calculate the power spectrum and bispectrum, we first seperate $u_\phi$ into two parts $u_\phi = u_\phi^{\rm vac} + u_\phi^{\rm s}$ where $u_\phi^{\rm vac}$ is the solution to the homogeneous part of \eqref{eomf} that reduces to Bunch-Davies vacuum on small scales, whereas $u_\phi^{\rm s}$ stands for the particular solution obtained by the Green's function. These two components are not statistically correlated, and therefore the dimensionless power spectrum can be written as,
\bea
\nn \fr{k^3}{2\pi^2}\langle \zeta_{\vec{k}}(\tau)\zeta_{\vec{k}'}(\tau)\rangle  &\equiv& \Big[\Delta_{\zeta,\rm v}^2 (k) + \Delta_{\zeta,\rm s}^2(k)\Big] \delta(\vec{k}+\vec{k}'),\\
&=&\left(\fr{\mathcal{H}}{\phi_{_0}' a}\right)^2 \Big[\langle  u_\phi^{\rm v} u_\phi^{\rm v} \rangle +\langle  u_\phi^{\rm s} u_\phi^{\rm s}\rangle\Big].
\eea
Using the solution to the homogenous part of \eqref{eomf}, 
one can obtain the standard vacuum contribution to the power spectrum. At leading order in slow-roll expansion, this gives,
\beq\label{pv}
\Delta_{\zeta,\rm v}^2 (k) = \fr{H^2}{8\pi^2\ep\Mp^2}.
\eeq
In this section, we are mainly interested in the sourced power spectrum. For this we denote the source term in the right hand side of \eqref{eomf} as $J_{\vec{k}}(\tau)$, and write the sourced two point correlator of $\zeta$ as

\beq\label{sp}
\langle \zeta_{\vec{k}}(\tau)\zeta_{\vec{k}'}(\tau)\rangle_{\rm s} = \left(\fr{\mathcal{H}}{\phi_{_0}'}\right)^2 \int{\d\tau'}\d\tau'' ~\fr{G^{3/2}_k(\tau,\tau')}{a(\tau)}\fr{G^{3/2}_k(\tau,\tau'')}{{a(\tau)}}~\langle~ J_{\vec{k}}(\tau')~J_{\vec{k'}}(\tau'') ~\rangle,
\eeq
where 
\beq\label{ss1}
J_{\vec{k}}(\tau') = - \fr{2 N_{\sgm} \sqrt{\ep\es}}{a(\tau')~f}\int\fr{\d^3p}{(2\pi)^{3/2}}~ A'_+(|\vec{k}-\vec{p}|,\tau')~|\vec{p}|~ A_+(|\vec{p}|,\tau')~ \hat{\mathcal{O}}_{i,\vec{k}-\vec{p}} ~\hat{\mathcal{O}}_{i,\vec{p}}.
\eeq
In writing the expression above, we have only considered the amplified gauge field mode functions and used the following decomposition of the electric and magnetic fields
\bea\label{EBD}
\nn E_i(\vec{k},\tau) = -\fr{1}{a(\tau)^2}~ A'_+(|\vec{k}|,\tau) ~\hat{\mathcal{O}}_{i,\vec{k}},\\
B_i(\vec{k},\tau) = \fr{1}{a(\tau)^2}~ |\vec{k}|~A_+(|\vec{k}|,\tau) ~\hat{\mathcal{O}}_{i,\vec{k}}
\eea
with $\hat{\mathcal{O}}_{i,\vec{k}}\equiv {\epsilon}^{+}_i(\vec{k})\left[\hat{a}(\vec{k})+\hat{a}^{\dagger}(-\vec{k})\right]$. Note that in the second line of \eqref{EBD}, we have also used the identity, $\epsilon_{ijk}k_j {\epsilon}^{+}_k(\vec{k})= -i k \epsilon^{+}_i(\vec{k})$. Making use of the solutions of the gauge field mode functions in \eqref{MEAS} (noting the relation in \eqref{mfst}) and the commutation relations in \eqref{cr}, it is straightforward to evaluate the correlator in \eqref{sp}, details of which we leave to the Appendix C,
\beq
\langle \zeta_{\vec{k}}(\tau)\zeta_{\vec{k}'}(\tau)\rangle_{\rm s} = \fr{2\pi^2}{k^3}~ 2.9 \times 10^{-4} ~\Big(\ep N_{\sgm}\Delta_{\zeta,\rm v}^2 (k) \Big)^2 ~\fr{e^{4\pi \xi}}{\xi^6}~ \delta(\vec{k}+\vec{k}'). 
\eeq
Therefore, combining with the vacuum contribution in \eqref{pv}, the total scalar power spectrum is given by
\beq\label{psz}
\Delta_{\zeta}^2 (k) = \Delta_{\zeta,\rm v}^2 (k)\left[1 + \gamma_s\Big(\ep N_{\sgm} \Big)^2 \Delta_{\zeta,\rm v}^2 (k) ~\fr{e^{4\pi \xi}}{\xi^6}\right],
\eeq
where $\gamma_s \equiv 2.9 \times 10^{-4}$. We now turn to the calculation of the bispectrum which is simply given by the equal time correlator of the curvature perturbation,
\beq
\langle \zeta_{\vec{k}_1}(\tau)\zeta_{\vec{k}_2}(\tau)\zeta_{\vec{k}_3}(\tau)\rangle = B_{\zeta}(k_1,k_2,k_3)~ \delta(\vec{k}_1+\vec{k}_2+\vec{k}_3), 
\eeq 
where $B_{\zeta}$ depends only on the magnitudes of the three external momenta. In general, bispectrum also includes contributions from the vacuum fluctuations similar to the power spectrum. However, this contribution is slow-roll supressed and in the presence of particle production in the gauge field sector, we expect the sourced term to dominate the bispectrum. We therefore write,
\begin{align}\label{sb}
\nn \langle \zeta_{\vec{k}_1}(\tau)\zeta_{\vec{k}_2}(\tau)\zeta_{\vec{k}_3}(\tau)\rangle &= -\left(\fr{\mathcal{H}}{\phi_{_0}'}\right)^3 \int\d\tau_1\d\tau_2\d\tau_3 ~\fr{G^{3/2}_k(\tau,\tau_1)}{a(\tau)}\fr{G^{3/2}_k(\tau,\tau_2)}{{a(\tau)}}\fr{G^{3/2}_k(\tau,\tau_3)}{{a(\tau)}}\\\nn\\
&~~~~~~~~~~~~~~~~~~~~~~~~~~~~~~~~~~~~~~~~~~~ \times~\langle~ J_{\vec{k}_1}(\tau_1)~J_{\vec{k}_2}(\tau_2)J_{\vec{k}_3}(\tau_3) ~\rangle.
\end{align}
The calculation of the bispectrum is a bit more involved compared to the power spectrum. We refer interested readers to the Appendix C. It is hard to evaluate the bispectrum for a general shape triangle that is formed by the three external momenta in \eqref{sb}. We therefore spell out the result in the equilateral configuration which nevertheless is expected to dominate the bispectrum \cite{Barnaby:2010vf}:
\beq\label{bsptrm}
B_{\zeta}(k,k,k) \simeq  10^{-9}~ N_{\sgm}^3~ \left(\fr{H}{\Mp}\right)^{6} \fr{e^{6\pi \xi}}{\xi^9} \fr{1}{k^6}.
\eeq
Note that scalar contribution to non-gaussianities we found here is much larger than the level of non-gaussianity induced by tensor perturbations \cite{Cook:2013xea}. To match with the CMB data, we are interested in the non-linearity parameter $\fnl$ which can be defined at the equilateral configuration as
\beq\label{fnl0}
\fnl^{\rm eq}(k,k,k) = \fr{10}{9} \fr{k^6}{(2\pi)^{5/2}} \fr{B_\zeta(k,k,k)}{[\Delta_\zeta^2(k)]^2}.
\eeq   
Finally, combining \eqref{bsptrm} and \eqref{fnl0} gives
\beq\label{fnl}
\fnl^{\rm eq} \simeq 5.4 \times 10^{-6} ~\fr{\Big(\ep~ N_{\sgm}~\Delta_{\zeta,\rm v}^2 (k) \Big)^3}{[\Delta_\zeta^2(k)]^2} ~ \fr{e^{6\pi \xi}}{\xi^9}.
\eeq

Now, we would like to comment on our results in the light of the recent investigations in the literature. The expressions we found for the scalar power spectrum and the non-linearity parameter shows the extent of the leak of the non-gaussianities from the hidden sector to the visible sector through the process: $\delta A + \delta A \to \delta\sgm \to \delta\phi$. In comparison with the case that the visible sector (inflaton) directly couples to the gauge field \cite{Barnaby:2010vf}, \ie through the process $\delta A + \delta A \to \delta\phi$, the non-gaussian contribution we present here is reduced by factors of $\ep N_{k}^{\sgm}$ in \eqref{psz} and \eqref{fnl}. On the other hand, we see that even if we set $N_{\sgm} =1$, the sourcing of the curvature perturbation in this channel is two orders of magnitude larger than those considered in \cite{Barnaby:2012xt} where only the contributions to the curvature correlators that are induced through gravity are taken into account (\eg compare the equations (108) and (110) in that work with \eqref{psz} and \eqref{fnl} in here). We note that the importance of the process, $\delta A + \delta A \to \delta\sgm \to \delta\phi$, was first considered in \cite{Ferreira:2014zia}, then in \cite{Namba:2015gja} and our results in this section are in qualitative agreement with the ones presented in those papers.    
\subsection{Tensors}\label{T}
In this section, we briefly review the calculation of the sourced tensor power spectrum. For tensor bispectrum signatures and their implications, see \eg \cite{Cook:2013xea}. To study the effects of particle production in the tensor sector, it is enough to consider the following action\footnote{See for example, the equations \eqref{SG} and \eqref{haa} in the Appendix A} 
\beq
S_{\rm GW} = \int \d^4 x \Bigg[\fr{\Mp^2 a^2}{8}\Big(h'_{ij}h'_{ij}-\partial_k h_{ij}\partial_k h_{ij}\Big)-\fr{a^4}{2}h_{ij}\Big(E_iE_j+B_iB_j\Big)\Bigg],
\eeq
where $h_{ij}$ is transverse, $\partial_i h_{ij}=0$ and traceless $h_{ii}=0$. We can decompose $h_{ij}$ in the usual way to obtain the equation of motion of the canonically normalized field describing tensor modes, 
\beq\label{hdec}
h_{ij}(\vec{x},\tau) = \fr{2}{a(\tau) \Mp } \int \fr{\d^3 k}{(2\pi)^{3/2}} e^{i\vec{k}.\vec{x}} \sum_{\lambda=\pm} \Pi^{*}_{ij,\lambda}~ h_\lambda (\vec{k},\tau), 
\eeq
where the polarization operators are 
\beq
\Pi^{*}_{ij,\pm} = \epsilon^{\pm}_i(\vec{k}) \epsilon^{\pm}_j(\vec{k}). 
\eeq
Plugging the decomposition \eqref{hdec} inside the action, we obtain the following equation of motion for $h_\lambda (\vec{k},\tau)$,   
\beq\label{meh}
\left(\partial_\tau^2 +k^2 -\fr{2}{\tau^2}\right) h_\lambda (\vec{k},\tau) = J_\lambda (\vec{k},\tau), 
\eeq
where the source term can be written as a convolution in terms of the electric and magnetic fields
\beq
J_\lambda (\vec{k},\tau)=-\fr{a^3}{\Mp} \Pi_{ij,\lambda} \int \fr{\d^3 p}{(2\pi)^{3/2}} \left[E_i(\vec{k}-\vec{p})E_j(\vec{p})-B_i(\vec{k}-\vec{p})B_j(\vec{p})\right].
\eeq
The solution to the equation \eqref{meh} can be written as a sum of vacuum mode and the sourced mode that are uncorrelated: $h_\lambda = h_\lambda^{\rm v} + h_\lambda^{\rm s}$ where the formal solution for sourced term is given by
\beq
h_\lambda^{\rm s}(\vec{k},\tau) = \int d\tau' G_k^{3/2}(\tau,\tau') J_\lambda(\vec{k},\tau').
\eeq
Using the standard canonical quantization scheme for the gauge fields and the vacuum mode, the tensor power spectrum can be calculated via 
\beq
\fr{k^3}{2\pi^2}\sum_\lambda \langle h_\lambda h_\lambda \rangle = \Delta_h^2(k) \delta(\vec{k}+\vec{k}').
\eeq
This calculation appeared in many places before
\cite{Barnaby:2010vf,Sorbo:2011rz,Cook:2013xea}, therefore we will simply state the final result 
\beq\label{tps}
\Delta_h^2(k) = 16 \ep~ \Delta_{\zeta,\rm v}^2 (k) \left[1 + \gamma_t~\ep ~\Delta_{\zeta,\rm v}^2 (k) ~\fr{e^{4\pi \xi}}{\xi^6}\right].
\eeq 
Here, the numerical factor is $\gamma_t \equiv 3.4 \times 10^{-5}$ and we have described the tensor power spectrum in terms of the power spectrum of the scalar vacuum fluctuations, $\Delta_{\zeta,\rm v}^2$   for convenience.
\section{Phenomenology}\label{Sec4}
In the previous section, we have identified the key observables in terms of the parameter set $\{\xi, \ep, N_{\sgm} \}$ of the model. We now compare them with cosmological observations in order to constrain this parameter space. We will see that bounds on the non-gaussianities from the CMB data, together with the requirement that the sourced tensor modes dominate over the vacuum fluctuations of the metric leads to certain requirements for the background dynamics of the model. 

\subsection{Normalization of the power spectrum}
The recent Planck data provide the amplitude of the power spectum at the pivot scale $k_* = 0.05 ~{\rm Mpc}^{-1}$, $\Delta^2_\zeta(k_*) \equiv A_s = 2.2 \times 10^{-9}$ \cite{Ade:2015lrj} which applies to the total power spectrum we have found in the previous section. In the $\Delta_{\zeta,\rm v} - \xi$ plane, the normalization of the power spectrum is satisfied along the curve as a function of $\ep$ and $N_{\sgm}$, 
\beq\label{pcurve}
\Delta^2_{\zeta,\rm v} (\xi,\ep, N_\sgm)= \fr{e^{-4\pi\xi}}{2f_{\gamma_s}(\xi,\ep, N_{\sgm})}\Big[-1+\sqrt{1+4A_sf_{\gamma_s}(\xi,\ep, N_{\sgm})e^{4\pi\xi}}~\Big] 
\eeq
where we have defined $f_{\gamma_s} \equiv \gamma_s (\ep N_\sgm)^2 /\xi^6$. It is clear that if the second term inside the square root is smaller than unity, we recover the standard result: $\Delta^2_{\zeta,\rm v} = A_s$. The value of $\xi$ where the sourced contribution is comparable to the vacuum depends on the specific model (through $\ep$ at horizon crossing) of inflation and the number of e-folds the spectator field rolls, $N_{\sgm}$, via the relation
\beq 
\fr{e^{4\pi\xi}}{\xi^6} = \fr{1}{4 A_s \gamma_s}(\ep N_{\sgm})^{-2}.
\eeq
From \eqref{pcurve}, we see that at fixed $\ep$ and $N_{\sgm}$, as $\xi$ is increased, source contribution starts to dominate and $\Delta^2_{\zeta,\rm v}$ needs to be exponentially decreased to avoid over production of scalar fluctuations. The real concern here is to keep the sourced contribution sub-dominant compared to the standard contribution of vacuum fluctuations. As we will show in the next section, the constraints from non-gaussianity dominate the parameter space of interest, making the sourced contribution in the scalar power spectrum small in \eqref{psz}.
\subsection{Constraints on non-gaussianity} 

\begin{figure}[t!]
\begin{center}
\includegraphics[scale=0.62]{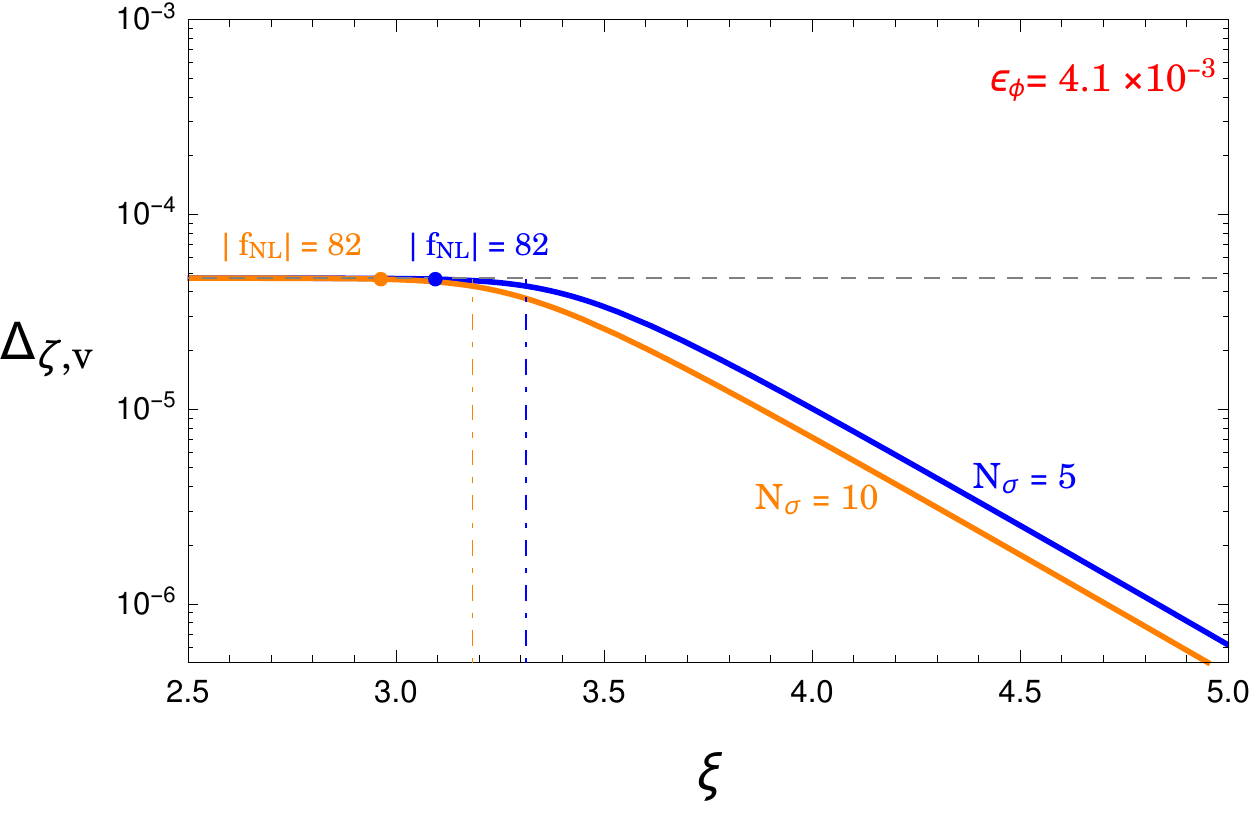}\includegraphics[scale=0.62]{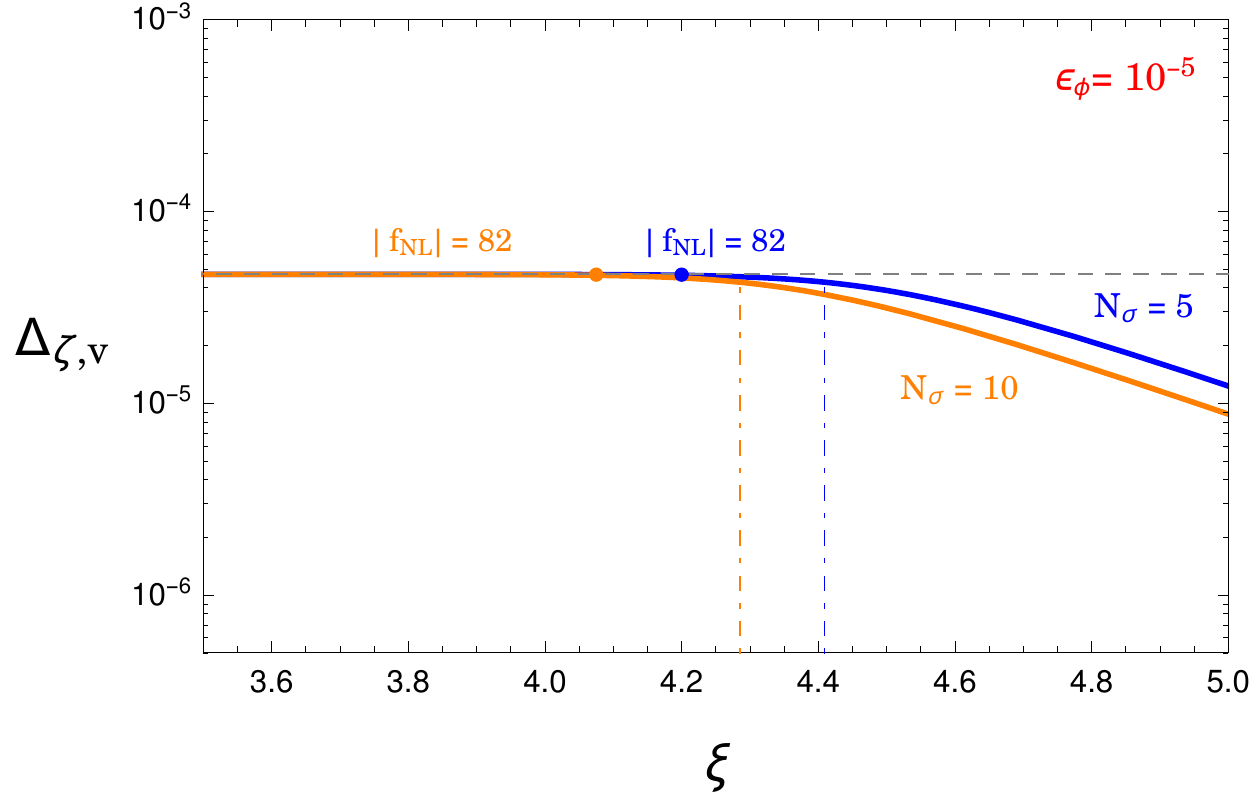}
\end{center}
\caption{The plane of $\Delta_{\zeta,\rm v} - \xi$ that leads to the observed power spectrum normalization for $N_{\sgm} =\{5,10\}$, $\ep = 4.1\times 10^{-3}$ ($\ep = 10^{-5}$) on the left (right), and the respective bounds $|f^{\rm eq}_{\rm NL}|=82$ on this curve. Dotted dashed lines indicates the value of $\xi$ where sourced power spectrum becomes comparable to the vacuum contribution and the grey dashed line stands for the reference value of $A_s = 2.2 \times 10^{-9}$.\label{fig:A}}
\end{figure}

CMB observations provide strict constraints on the 3-pt correlators of the curvature perturbation $\zeta$. In the model under consideration, the signal peaks at the equilateral shape\footnote{In the single field version of this model, the overlap between the signal and the equilateral template is shown to be $93 \%$ \cite{Barnaby:2011vw}. This tells us that there is approximately a $10 \%$ uncertainty in constraining the model. We expect to have a similar situation here, as the nature of particle production is essentially the same as in \cite{Barnaby:2011vw}.} and therefore the following bounds from CMB data can be applied: $\fnl^{\rm eq} < -16 \pm 70$ (temperature only), $\fnl^{\rm eq} < -4 \pm 43$ (T+E) at $68 \%~ \rm CL$ \cite{Ade:2015ava}. In the following, we will impose $\fnl^{\rm eq}$ in \eqref{fnl} to be smaller than $2\sgm$ limits published by Planck in order to constrain the parameter space of the model, 
\beq\label{bfnl}
|\fnl^{\rm eq}| < 124 ~~({\rm T}),~~~~~~~~ |\fnl^{\rm eq}| < 82 ~~(\rm T+E).
\eeq
First, we simplify our notation by introducing the quantity $X \equiv \ep~ e^{2\pi\xi}/\xi^3$ and relabeling the vacuum power spectrum as $\Delta^2_{\zeta,\rm v}\equiv \mathcal{P}$, following \cite{Barnaby:2012xt,Ferreira:2014zia}. We rewrite the expression in \eqref{fnl} as a function of these parameters and impose the bound $|\fnl^{\rm eq}| < 82$, which turns into a constraint for the following combination of the model parameters   
\beq\label{ngb}
 (N_{\sgm})^2 ~\mP^2 X^2  < 1.8 \times 10^{-7} \left(\fr{\fnl^{\rm eq}}{82}\right)^{2/3},
\eeq
where we have used the power spectrum normalization $\Delta^2_{\zeta}(k_*) \equiv 2.2 \times 10^{-9}$. Now, using the bound \eqref{ngb} on the model parameters, we see that the sourced contribution of the scalar power spectrum \eqref{psz} is sub-dominant compared to the vacuum contribution, 
\beq
\nn \Delta_{\zeta,\rm s}^2 < 5.1 \times 10^{-11} \left(\fr{\fnl^{\rm eq}}{82}\right)^{2/3}\ll A_s. 
\eeq
This result implies that in light of the constraints from non-gaussianity, we can simply assume that the power spectrum is dominated by the vacuum contribution and take $\Delta_{\zeta}^2 = \Delta_{\zeta,\rm v}^2 \equiv \mP = A_s$. One can reach the same conclusion from Figure \ref{fig:A} where the curve in \eqref{pcurve} is shown for different values of $N_\sgm$ and $\ep$ along with the reference value $\fnl^{\rm eq} = 82$ for each case. As $\fnl^{\rm eq} \propto N_\sgm^3$, the bound on non-gaussianity saturates at different points on this curve. In Figure \ref{fig:A}, we also show the value of $\xi$ where the sourced power spectrum becomes comparable to the vacuum contribution with dotted dashed lines. It is clearly visible from the figure that the bound on non-gaussianity always saturates at smaller $\xi$ compared to the value of $\xi$ where the sourced scalar power spectrum becomes comparable to the vacuum one. Another difference between the right and left panel of Figure \ref{fig:A} is that for smaller $\ep$, both the value of $\xi$ where the sourced power spectrum becomes comparable to the vacuum one and the value where the non-gaussianity bound is saturated shifts towards larger values of $\xi$.

In summary, we would like to emphasize that the constraints on non-gaussianity from Planck implies that the sourced contribution of the power spectrum is small and no independent constraints on the parameter space $\{\xi, \ep, N_{\sgm}\}$ of the model arise from the normalization of the power spectrum. Therefore, in  determining the strength of the secondary contribution to the tensor power-spectrum in \eqref{tps}, we can simply apply the bounds shown in \eqref{bfnl} using \eqref{fnl} and utilize the standard form for the scalar power spectrum, $\Delta^2_{\zeta}=\Delta^2_{\zeta,\rm v}\equiv\mP$. In the next section, we will examine the strength of the sourced GW's in the light of the non-gaussianity bounds and by taking into account restrictions that might arise from successful model building. 
\subsection{Limits on the sourced tensor power spectrum}
Following our discussion in the previous section, tensor to scalar ratio can be parametrized as
\beq\label{r}
r  \simeq  16 \ep~  \left[1 + \gamma_t~\ep ~\mP ~\fr{e^{4\pi \xi}}{\xi^6}\right] \equiv r_{\rm v}\left[1 + \delta_t\right],
\eeq
where we defined the excess power in tensors by $\delta_t \equiv r_{\rm s}/r_{\rm v}$ with $r_{\rm v} = 16 \ep$. Using the expression for the scalar vacuum power spectrum in \eqref{pv}:  $\ep \Delta^2_{\zeta,\rm v}\equiv \ep \mP = (1/8\pi^2)(H^2/\Mp^2)$, we re-write the excess power $\delta_t$ as
\beq\label{dt1}
\delta_t (\xi,H) =  4.3 \times 10^{-7} \left(\fr{H}{\Mp}\right)^2 \fr{e^{4\pi \xi}}{\xi^6}.
\eeq
We find writing $\delta_t$ in this way is more convenient since it can be solely expressed in terms of the physical Hubble scale $H$ and the parameter $\xi$ that controls the strength of the particle production. Similarly, we can re-write the non-linearity parameter in \eqref{fnl} as 
\beq\label{bfnl1}
\left(\fr{\fnl^{\rm eq}(\xi,H)}{82}\right)^{2/3} \simeq 9.1 \times 10^2 ~ (N_{\sgm})^2 \left(\fr{H}{\Mp}\right)^4 \fr{e^{4\pi \xi}}{\xi^6}.
\eeq
The expression above essentially restricts the paramater $\xi$ that controls the efficiency of particle production (through the factor $e^{4\pi \xi}/\xi^6$) as a function of Hubble rate $H$ and $N_{\sgm}$, once the bound on $|f^{\rm eq}_{\rm NL}|$ is applied. Combining \eqref{dt1} with \eqref{bfnl1} and imposing \eqref{bfnl}, we obtain the following upper bound on the excess power in tensors,
\beq\label{dtf}
\boxed{
\delta_t \lesssim 28 ~\left(\fr{10}{N_{\sgm}}\right)^2 \left(\fr{|\fnl^{\rm eq}|}{82}\right)^{2/3} \left(\fr{10^{12}~ {\rm GeV}}{H}\right)^2 .}
\eeq
We see that once the limits from non-gaussianity are respected, we can still accomadate large sourced contribution with respect to the vacuum  GW's for arbitrarily small values of Hubble rate $H$ during inflation, corresponding to small inflationary energy scales $\rho_{\rm inf} = 3H^2\Mp^2$. However, the allowed parameter space can be constrained further by taking into account the back-reaction constraints (\eg see Section \ref{Sec2}) that we investigate now.

{\bf Back-reaction constraints.}
In this model, the particle production proceeds at the expense of the kinetic energy of the pseudoscalar $\sgm_{_0}(t)$ and therefore we require the kinetic energy of the scalar to be larger than the energy density contained in the gauge fields, $\dot{\sgm}^2_{_0}/2 > \rho_A$ (See eq. \eqref{EV}), which in turn can be considered as the source of the secondary tensor modes. The condition, $\dot{\sgm}^2_{_0}/2 > \rho_A$ can be re-written as an independent bound on $\xi$ as a function of the ratio of the slow-roll parameters (kinetic energies) of background fields
\beq\label{br1}
\fr{e^{4\pi \xi}}{\xi^6} < 1.7 \times 10^{21} \left(\fr{\es}{\ep}\right)^2 .
\eeq   
Plugging this expression in \eqref{dt1}, we reach at the following bound
\beq\label{dt2}
\boxed{
\delta_t < 1.2 \times 10^2 \left(\fr{\es}{\ep}\right)^2 \left(\fr{H}{10^{12}~ {\rm GeV}}\right)^2.}
\eeq
We see that at fixed Hubble scale $H$ (or $\rho_{\rm inf}=3H^2\Mp^2$), it is easier to make the sourced contribution larger by increasing the ratio $\es/\ep$ as there is enough kinetic energy available in $\sgm$ compared to the inflaton. On the other hand, for fixed ratio of the kinetic energies of the scalar fields $\es/\ep$, reducing $H$ will make the bound stronger as there is less total energy available in the system. Note that in \cite{Ozsoy:2014sba}, the condition $\es<\ep$ is imposed. However as shown in \cite{Peloso:2016gqs} recently there is no reason to enforce this condition: at the level of background, having the condition $\es > \ep$ does not affect the quasi de-Sitter expansion as far as $\ep \ll 1$ and $\es \ll 1$, nor the Friedmann equation as we assume $\rho_\sgm \ll V(\phi)$. The only potential affect of $\es > \ep$ is in the spectral tilt of the scalar perturbations which receives contributions proportional to $\es$,
\beq\label{si}
n_s -1 \simeq 2\eta_\phi -6\ep - 4\es \approx 2\eta_\phi - 4\es,
\eeq 
where we assumed $\es > \ep$ in the last equality. Taking into account the observed value of the tilt, $|n_s -1|\simeq 0.03$ \cite{Ade:2015lrj}, one may only require $\es < 10^{-2}$ to avoid fine tunings in \eqref{si} in which case the scalar spectral tilt will be controlled by $\eta_\phi$. Therefore as long as $\es < 10^{-2}$, for the $N_{\sgm}$ e-folds in which $\sgm$ is rolling, the condition $\es>\ep$ has no crucial phenomenological implication.
 
\begin{figure}[t!]
\begin{center}
\includegraphics[scale=0.72]{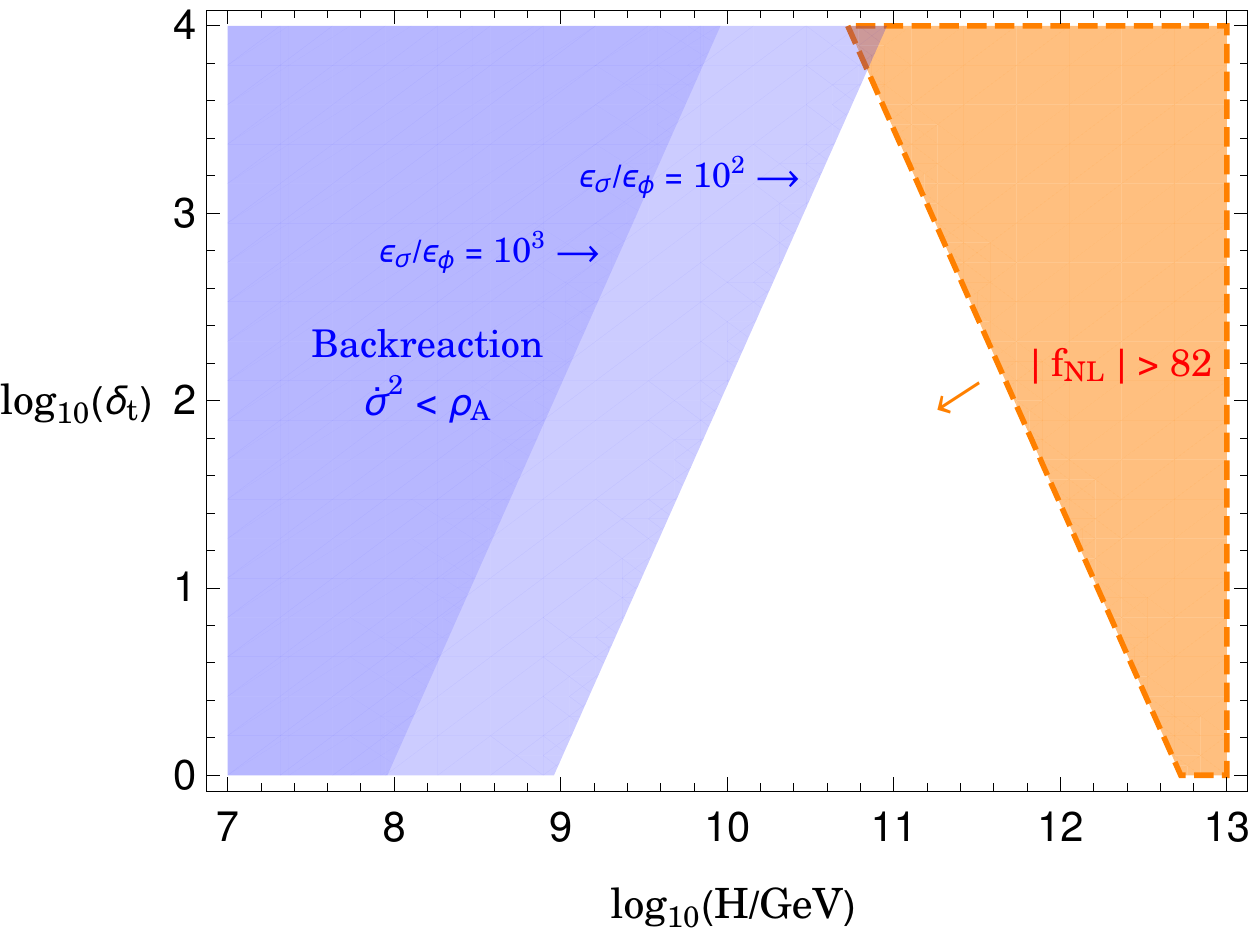}
\end{center}
\caption{Constraints from non-gaussianities and back-reaction on the excess power of tensors are shown in the $\delta_t -H$ plane. The back-reaction constraint is shown in blue shades (\eg \eqref{dt2}) where we choose $\es/\ep =\{10^2, 10^3\}$ for illustrative purposes. It is clearly visible that the allowed region (white) expands as a function of $\es/\ep$. On the other hand, the boundry for the non-gaussianity constraints (shown in orange shades) is obtained by assuming the value $N_{\sgm} = 10$ in \eqref{dt1}. \label{fig:B}}
\end{figure}

Following the discussion above, in Figure \ref{fig:B}, we show the constraints obtained from \eqref{dtf} and \eqref{dt2} in the $\delta_t - H$ plane. In this plot, we set $N_{\sgm}=10$ in \eqref{dtf} to present the constraints from non-gaussianities. We see that even if the spectator scalar $\sgm$ rolls for ten e-folds after the CMB modes exit the horizon, the model can still account for a sourced tensor power spectrum that is larger than the vacuum contribution in the large portion of its parameter space. Especially, we see that the strength of the sourced signal can be made much larger for small values of Hubble scale $H$ corresponding to lower energy scales for inflation, $\rho_{\rm inf} = 3H^2\Mp^2$. Future bounds on non-gaussianity will push the allowed $\delta_t$ to lower values while restricting the maximum allowed $H$ as indicated by the orange arrow in Figure \ref{fig:B}.

{\bf Tensor to scalar ratio.}
In the model under consideration, the rolling spectator $\sgm$ amplifies only one of the helicity modes of the gauge field, which in turn leads to chiral GW's through the mechanism $\delta A + \delta A \to \delta g$ \cite{Sorbo:2011rz}. It is therefore important to check the strength of the signal since chiral GW's are a potential distinguishing feature of the model whenever the sourced contribution dominates. As shown in  \cite{Cook:2013xea,Gluscevic:2010vv}, a detection of chiral gravitational waves is possible, in principle by experiments such as Spider \cite{Crill:2008rd} and CMBpol \cite{Baumann:2008aq}.

\begin{figure}[t!]
\begin{center}
\includegraphics[scale=0.6]{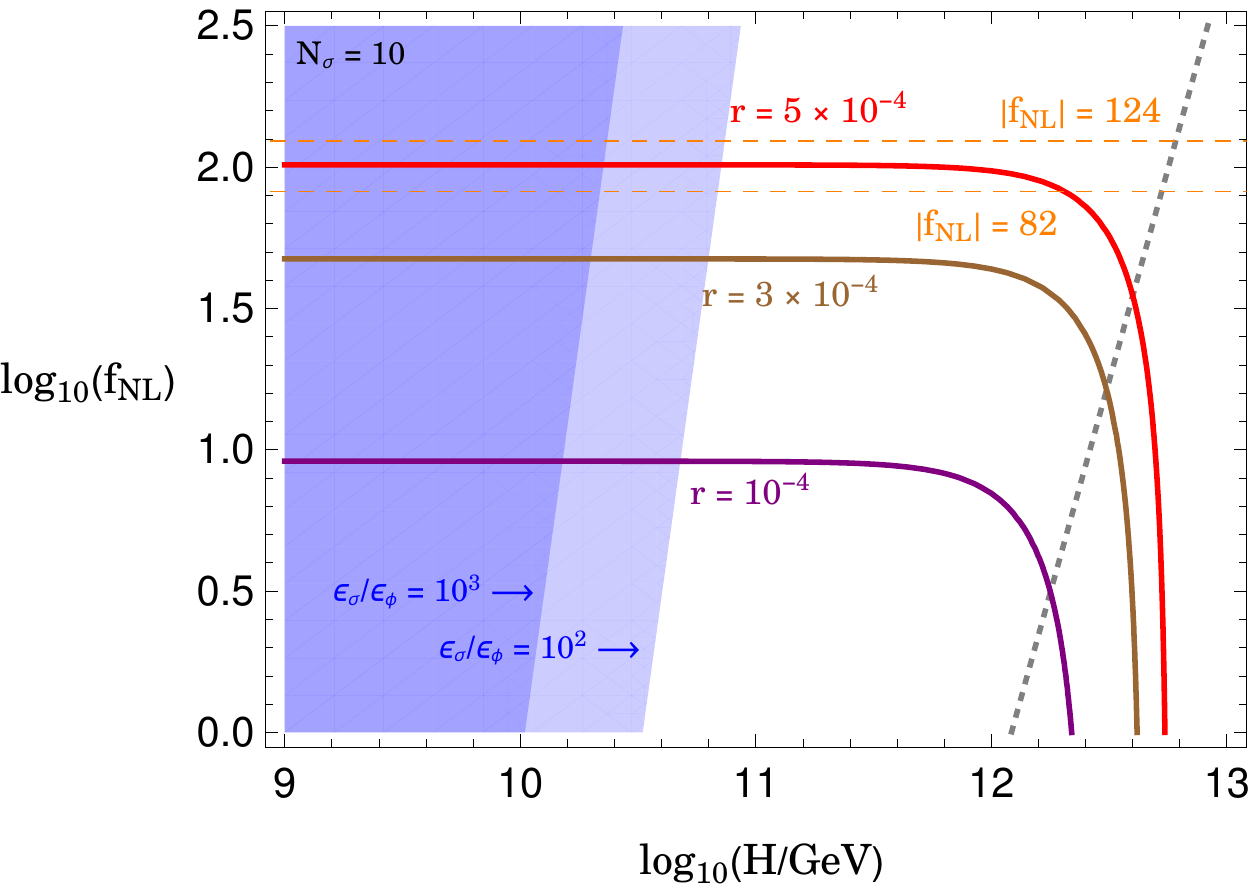}~\includegraphics[scale=0.6]{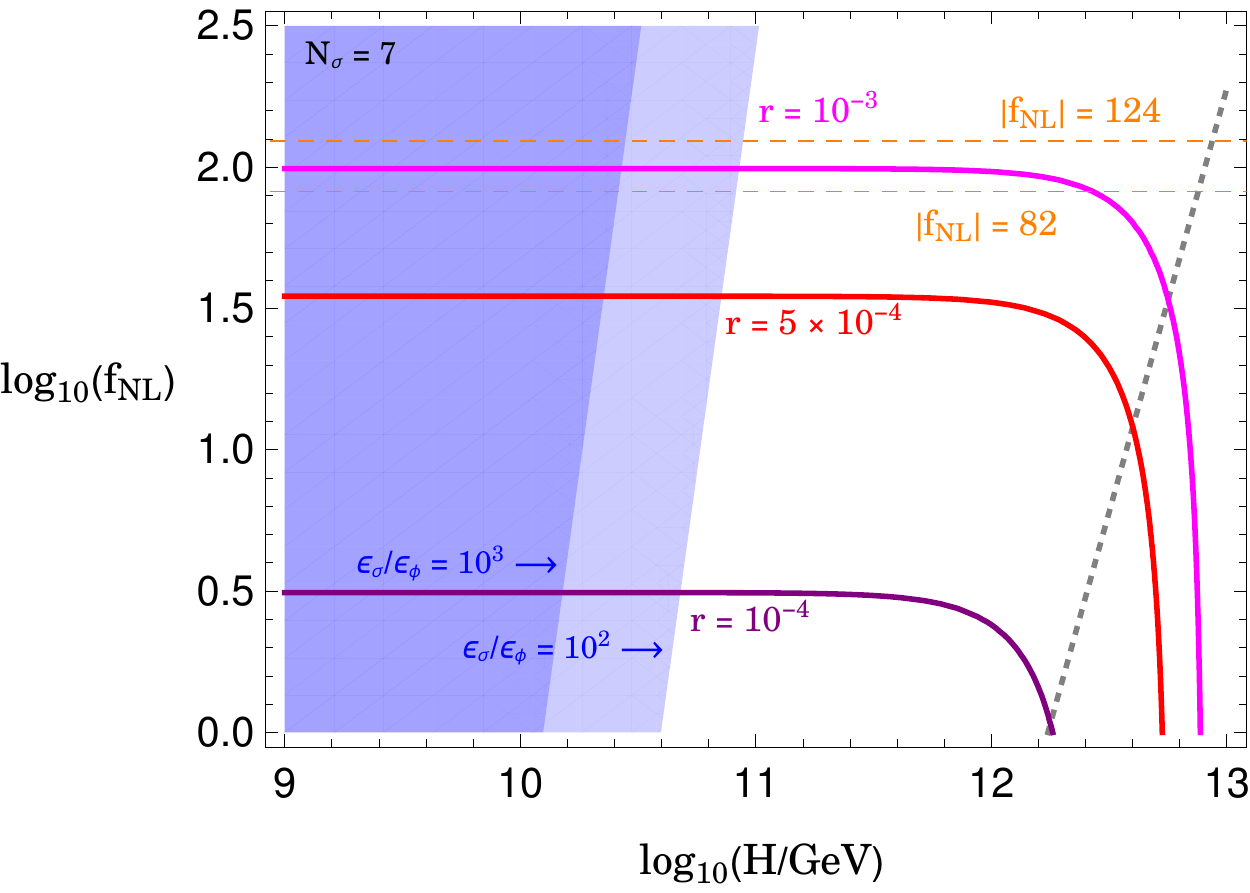}
\end{center}
\caption{Solid contour lines: tensor to scalar ratio \eqref{r1} in the $H-\fnl^{\rm eq}$ plane. The back-reaction constraint is shown in blue shades similar to the Figure \ref{fig:B}. The dashed orange lines indicates the current $2\sgm$ limits on $\fnl^{\rm eq}$ \eqref{bfnl} where the region above these lines are excluded. The left hand side of the dotted gray line indicates the parameter space where the sourced contribution dominates over the vacuum fluctuations of the metric, \ie $\delta_t >1$. \label{fig:C}}
\end{figure}

We will present our results on tensor to scalar ratio in the $H-\fnl^{\rm eq}$ plane in order to guide the eye for the current and future bounds on non-gaussianity. In terms of the parameter set $\{H,\fnl^{\rm eq},N_\sgm\}$, tensor to scalar ratio is given by the following expression
\beq\label{r1}
r\simeq 1.6 \times 10^{-5} \left(\fr{H}{10^{12}~ {\rm GeV}}\right)^2 + 5.7 \times 10^{-4} \left(\fr{10}{N_{\sgm}}\right)^2 \left(\fr{|\fnl^{\rm eq}|}{124}\right)^{2/3},
\eeq
where we rewrote the vacuum contribution $r_{\rm v}= 16\ep$ using the normalization of the power spectrum: $\ep = (8\pi^2 A_s)^{-1}(H^2/\Mp^2)$. From this expression we can immediately read the maximum strength of the sourced contribution (second term) to $r$ which can be as large as $5.7 \times 10^{-4}$ for $N_\sgm =10$. Note also that to make the source contribution dominant, the following condition needs to be satisfied in the $H-\fnl^{\rm eq}$ plane 
\beq
\left(\fr{|\fnl^{\rm eq}|}{124}\right)^{2/3} \gtrsim 2.8~ \left(\fr{N_{\sgm}}{10}\right)^2 \left(\fr{H}{10^{13}~ {\rm GeV}}\right)^2.
\eeq 
On the other hand, to show the back-reaction constraint in the $H-\fnl^{\rm eq}$ plane, we combine the equation \eqref{br1} with \eqref{bfnl1}, which allow us to replace the back-reaction constraint in \eqref{br1} by the following bound
\beq\label{br2}
\left(\fr{|\fnl^{\rm eq}|}{124}\right)^{2/3} \lesssim 3.3~ \left(\fr{N_{\sgm}}{10}\right)^2~\left(\fr{\es}{\ep}\right)^2 ~ \left(\fr{H}{10^{12}~ {\rm GeV}}\right)^4.
\eeq
In Figure \ref{fig:C}, we collect our results in \eqref{r1}-\eqref{br2} to show various contour lines of tensor to scalar ratio $r$ together with the constraints from back-reaction and on $\fnl^{\rm eq}$ in \eqref{bfnl}. We show our results in the $H-\fnl^{\rm eq}$ plane for two different realization of mult-field inflation where the spectator field $\sgm$ rolls for $N_\sgm = 10$ (left panel) and $N_\sgm = 7$\footnote{For $N_\sgm \leq 5$, the constant roll approximation ($\xi \approx constant.$) we have made in our calculations will become less reliable.}(right panel). In these plots, we parametrized the backreaction constraints in \eqref{br2} in terms of the ratio of kinetic energies of the background fields, similar to the case in Figure \ref{fig:B}. From the position of contour lines of $r$ in Figure \ref{fig:C}, we see that a tensor to scalar ratio as large as $r \sim 10^{-3}$ can be obtained, for which sourced GW's by the gauge fields dominate over the vacumm ones. In particular, considering the $2\sgm$ limits on the non-linearity parameter for the temperature data only $|\fnl^{\rm eq}| < 124$, a value of $r = 5\times 10^{-4}$ is still allowed while the spectator can roll for ten e-folds, $N_\sgm =10$, after the relevant modes exit the horizon. If we restrict the background model of the spectator by reducing the amount of e-folds to $N_\sgm =7$, even a value of $r = 10^{-3}$ is viable. Including the Planck polarization data $|\fnl^{\rm eq}| < 82$, restricts the allowed tensor to scalar ratio to smaller values for each case: $r \approx 3-4 \times 10^{-4}$ ($N_\sgm = 10$) and $r \approx 5-7 \times 10^{-4}$ ($N_\sgm = 7$). We see that although the signal decrease for larger values of $N_{\sgm}$, it is not too far below the projected sensitivity of the next stage CMB experiments, $r \simeq 10^{-3}-10^{-4}$ \cite{Abazajian:2016yjj, Finelli:2016cyd,Andre:2013afa}.    

{\bf Comments on the background model.} If we keep the original parametrization for the excess power in tensor modes \eqref{r} in terms of the slow-roll parameter of the inflaton $\ep$, we can gain some insight on the background model that allows for larger than vacuum sourced contribution:     
\beq
\delta_t \equiv \fr{r_{\rm s}}{r_{\rm v}} \simeq \fr{3.6 \times 10^{-3}}{\ep N_{\sgm}^2} \left(\fr{|\fnl^{\rm eq}|}{124}\right)^{2/3}.
\eeq
 
It is clear from this expression is that the sourced contribution to the tensor power can be made dominant only if the condition $\ep~ N_{\sgm}^2 < 3.6 \times 10^{-3}$ is satisfied. This condition have important implications for the background model of both inflaton and the spectator sector. For example, for a given model of inflation (by fixing $\ep$ at horizon crossing), it can be considered as an upper bound on the number of e-folds $N_{\sgm}$ that the spectator allowed to roll while keeping the sourced tensor contribution dominant compared to the vacuum one. We see that only for sufficiently flat potentials around horizon crossing, the spectator $\sgm$ is allowed to roll considerably. This can be also verified by the Figure \ref{fig:B}, where the allowed parameter space corresponds to relatively small values of the Hubble parameter $H$ which in fact corresponds to small values of $\ep$. Keep in mind that, one can not arbitrarily lower the Hubble scale $H$ to obtain a larger $N_\sgm$ both because the back-reaction constraints put a lower bound on $H$ and because for large $N_\sgm$ the observable signal $r$ in \eqref{r1} reduces considerably. We note that our discussion here is in agreement with the results presented in \cite{Namba:2015gja,Peloso:2016gqs} (see, for example, Figure 3 of \cite{Namba:2015gja}) where the authors considered only a transient roll\footnote{In the case of a transient roll for a few e-folds, the sourced tensor and scalar correlators obtain a scale dependence during which the signal peaks. This case has to be contrasted with the scale invariant signal that arise in the constant roll case we are considering here.} for the spectator $\sgm$.

\section{Conclusions}\label{Sec5}

In this paper, we have considered whether continuos particle production from a rolling spectator pseudoscalar (apart from the inflaton) can lead to a competive source of primordial gravity waves through its coupling to the Abelian gauge fields during inflation. For this purpose, we have identified the dominant contribution to the curvature perturbation in the presence of gauge field sources and showed that leading constraints on the parameter space of the model is due to the non-gaussian contribution to the scalar correlators that arise through the process $\delta A + \delta A \to \delta \sgm \to \delta\phi$, in agreement with the recent results \cite{Ferreira:2014zia, Namba:2015gja} (See also \cite{Caprini:2017vnn} for a discussion in a similar model).

Although the paramater space of the model is constrained when we apply bounds from non-gaussianity and backreaction effects, we find that it is still possible to generate observably large secondary GW's in this model (see Figure \ref{fig:B} and \ref{fig:C}). This conclusion is subject to two conditions, \ie in order to avoid large non-gaussian scalar fluctuations that invalidates a large sourced tensor power spectrum, we require
\begin{itemize}
\item The spectator scalar has to decay after a certain number of e-folds $N_{\sgm}$ after the observable scales exits the horizon,
\item At horizon crossing, the condition $\ep (N_{\sgm})^2 \lesssim 10^{-3}$ needs to be satisfied.
\end{itemize}
When these conditions are satisfied, we showed that one can achieve a tensor to scalar ratio as large as $r \approx 10^{-3}-10^{-4}$ for which the gauge field sources dominate over the vacuum fluctuations of the metric. In view of these results, we emphasize that the bound on the number of e-folds $N_{\sgm}$ that the spectator $\sgm$ roll depends on the slow-roll parameter at horizon crossing and thefore it is allowed to be larger for flatter potentials at horizon exit. As a consequence, we found that the multi-field model we considered in this work is still in good shape both from the observational and model building perspective: it can still give rise to observably large GW's that originates from the gauge field sources while the requirements on the background dynamics for this conclusion can be easily achieved within the multi-field slow-roll inflation.       
\section*{Acknowledgements}
It is a pleasure to thank Marco Peloso and Lorenzo Sorbo for useful comments, extensive discussions and suggestions on an earlier version of the manuscript. I would also like to thank Jayanth Neelakanta for useful discussions, Gianmassimo Tasinato and Ivonne Zavala for reading the manuscript and Scott Watson for encouragement. The work of OO is supported by the STFC grant ST/P00055X/1 and NASA Astrophysics Theory Grant NNH12ZDA001N.	 	 
\section*{Appendix A: ADM formalism}\label{AppA}
We focus on the action for the matter Lagrangian in \eqref{Lm} minimally coupled to Einstein gravity
\beq\label{SF}
S = \int \d^4 x \sqrt{-g}\left[\fr{\Mp^2}{2}R-\fr{1}{2}(\del\phi)^2 -\fr{1}{2}(\del\sigma)^2 - V(\phi,\sgm)-\fr{1}{4}F^2-\fr{\sgm}{8f}\fr{\eta^{\mu\nu\rho\sigma}}{\sqrt{-g}}F_{\mu\nu}F_{\rho\sigma}\right],
\eeq
and assume the full potential is seperable with no direct couplings between the fields, $V(\phi,\sgm) = V_\phi(\phi) + V_\sgm(\sgm)$. 
In order to study cosmological perturbations, we write the metric in the ADM form,
\beq
\d s^2 = -N^2 \d t^2 + \hat{g}_{ij} \left(\d x^{i}+N^{i}\d t\right)\left(\d x^{j}+N^{j}\d t\right),
\eeq
where $\hat{g}_{ij}$ is the spatial 3-metric defined on constant time surfaces. In this parametrization, the lapse $N$ and shift vector $N^{i}$ appear as Lagrange multipliers and hence can be integrated out from the action \eqref{SF}. 
To study the fluctuations around the inflationary background we consider the following gauge-fixing conditions
\bea\label{gauge}
\nn \phi(t,\vec{x}) &=& \phi_{_0}(t) + \delta \phi(t,\nn \vec{x}),\\
\nn \sgm(t,\vec{x}) &=& \sgm_{_0}(t) + \delta \sgm(t,\vec{x}),\\
\nn \hat{g}_{ij} &=& a^2(t) \left[\delta_{ij}+h_{ij}\right],\\
A_0 &=& \del_i A_i =0 
\eea
where $h_{ij}$ is transverse and traceless, \ie $h_{ii}=\del_ih_{ij}=0$.  We can now expand the action at a desired order to solve for $N$ and $N^{i}$ in terms of $\delta \phi, \delta \sgm, h_{ij}$ and $A_i$ perturbatively. Note that by our gauge choice above and due to the fact that gauge fields do not contribute to the background evolution, solutions to lapse function and the shift vector can not start at linear order in $A_i$ and $h_{ij}$. Therefore schematically we expect
\bea
N = N(\delta \phi, \delta \sgm) + \mathcal{O}(h^2,A^2),\\
N^{i} = N^{i}(\delta \phi, \delta \sgm) + \mathcal{O}(h^2,A^2).
\eea
In addition, to obtain the action upto third order in fluctuations $\delta \phi, \delta \sgm, h_{ij}, A_i$, it is enough to solve the lapse and shift to first order in fluctuations \cite{Maldacena:2002vr}.
First, we begin by writing the gravity sector in the ADM form
\beq
\label{EH} S_{\rm g} = \frac 12 \Mp^2 \int \! \d^4 x \:
\sqrt{-g} \, R = \frac 12 \Mp^2 \int \! \d^4 x \:
\sqrt{\hat g} ~N \, \big[ R^{(3)} + \frac{1}{N^2} (E^{ij} E_{ij} -
E^i{}_i {}^2) \big], \;
\eeq 
where $R^{(3)}$ is the 3 curvature associated with the spatial metric $\hat{g}_{ij}$ and $E_{ij}$ is related to the extrinsic curvature of constant time slices,
\bea
E_{ij} & \equiv & N K_{ij} = \fr{1}{2} [{\partial_t
{\hat g}}_{ij} - \hat \nabla_i N_j - \hat \nabla_j N_i] \; ,\\
R^{(3)} &=& \hat{g}^{ik}\del_l\Gamma^{l}_{ik} -\hat{g}^{ik}\del_k\Gamma^{l}_{il} + \hat{g}^{ik}\Gamma^{l}_{ik}\Gamma^{m}_{lm} -\hat{g}^{ik}\Gamma^{m}_{il}\Gamma^{l}_{km},
\eea
with the connection 
\beq
\Gamma^{k}_{ij} = \fr{1}{2} \hat{g}^{kl}\left(\del_i \hat{g}_{jl}+\del_j\hat{g}_{il}-\del_l\hat{g}_{ij}\right).
\eeq
Here, $\hat \nabla_i$ is the covariant derivative with respect to 3-metric $\hat{g}_{ij}$. Noting our gauge choice \eqref{gauge} for the spatial metric and its inverse $\hat{g}^{ij} = a^{-2} [\delta^{ij} - h^{ij}]$, up to second order in fluctuations we have 
\beq\label{SG}
S_{\rm g} =  \Mp^2 \int \! \d^4 x \:
a^3  \, \Bigg[ 3H^2\delta N - 3H^2 \delta N^2 -2H\delta N \partial_i N^{i} +\fr{1}{8}\left(\dot{h}_{ij}\dot{h}_{ij}-\fr{\del_k h_{ij}\del_k h_{ij}}{a^2}\right) \Bigg]. \;
\eeq 
We can neglect the higher order terms in the gravity sector as they are slow-roll suppressed compared to the interactions of the form $\mathcal{O}(hAA)$ in the matter sector.

Next, we focus on the action in the scalar sector while ignoring pseudo-scalar coupling in \eqref{SF} for now. Since both scalar fields are minimally coupled, their action will have the same form. Therefore we will refer to both scalar fields collectively using the notation $\phi_a=\phi_{_{0a}}+\delta\phi_a$ where $a=1,2$ refers to the set of fields $\{\phi,\sgm\}$. Similarly, for the potentials we use $V_a$ where $\{V_1,V_2\} =\{V_\phi,V_\sgm\}$. At leading order in scalar fluctuations, the action for a minimally coupled scalar field $\delta\phi_a$ with a potential $V_a$ is given by
\bea
\label{SM} S_{a} = \fr{1}{2}\int \! \d^4 x \:
\nn a^3 \, &\Bigg\{& \left[\delta \dot{\phi}_a^2-\fr{\left(\del_i\delta \phi_a\right)^2}{a^2}\right]- V_a''(\phi_{_{0a}})\delta X^2 -2\dot{\phi}_{_{0a}}\delta \dot{\phi}_a\delta N\\
&+& \dot{\phi}_{_{0a}}^2\delta N^2 + 2\dot{\phi}_{_{0a}} N^{i}\del_i\delta \phi_a -2 V_a'(\phi_{_{0a}})\delta \phi_a\delta N \Bigg\},\\
\eea 
where no summation on index $a$ is implied. Considering the total action for scalar fluctuations, $S_g +\sum_a S_a $, we can obtain solutions for the Lagrange multipliers $\delta N$ and $N^{i}$ in terms of the scalar fields as,
\bea
\label{LS1}2H\Mp^2~ \delta N &=& \sum_a \dot{\phi}_{_{0a}} \delta \phi_a ,\\
\label{LS2}-2 H \Mp^2 ~\del_i N^{i} &=& \sum_a \left(\dot{\phi}_{_{0a}}\delta \dot{\phi}_a + V_a'(\phi_{_{0a}})\delta \phi_a\right) +\left(6H^2\Mp^2 -\sum_a \dot{\phi}_{_{0a}}^2\right)\delta N.
\eea
Plugging these solutions back in the actions \eqref{EH} and \eqref{SM}, we obtain the following second order action for scalar fluctuations,
\bea
S_2 &=&\fr{1}{2}\int \! \d^4 x \:
\nn a^3 \, \Bigg\{ \left[\delta \dot{\phi}_a^2-\fr{\left(\del_i\delta \phi_a\right)^2}{a^2}\right]-m_{ab}^2 \delta \phi_a\delta \phi_b\Bigg\},\\
m_{ab}^2 &=& V_{,ab} + \left(3-\fr{\sum_c \dot{\phi}_{_{0c}}^2}{2H^2\Mp^2}\right)\fr{\dot{\phi}_{_{0a}}\dot{\phi}_{_{0b}}}{\Mp^2}+ \fr{\dot{\phi}_{_{0a}} V_{,b}+\dot{\phi}_{_{0b}} V_{,a}}{H\Mp^2}.
\eea
where summation over $a,b$ is implied and $V_{,a}= \partial V/\partial\phi_a$.
Note that in deriving the expressions above we have repeatedly used the background equations of motion
\bea\label{BGX}
\ddot{\phi}_{_{0a}} + 3H\dot{\phi}_{_{0a}} &+& V_a'(\phi_{_{0a}})=0, \\
-2 \dot{H}\Mp^2 &=& \sum_a \dot{\phi}_{_{0a}}^2 .
\eea 

As far as the dynamics of scalar fluctuations concerned, it is enough to consider the leading action above as higher order interactions will be slow-roll suppressed through the derivative of potentials $V^{(n)}_\phi,U^{(n)}_\sgm$ with $n\geq 3$. Particularly, in the presence of particle production in the gauge field sector, leading interactions will be due to direct couplings to gauge fields $A_i$ or due to the ones induced by gravity, \ie terms of the form $\delta g A A$. Our aim in the next section is therefore to focus on these interactions that arise from the last two terms in the action \eqref{SF}. 
\subsection*{Gauge Field Sector}
We are interested in the part of the action \eqref{SF} that includes gauge field and its interaction with the spectator sector $\sgm$ and gravitational fluctuations $\delta N, N^{i}$ and $h_{ij}$,
\beq
S = \int \d^4 x \sqrt{-g}\left[-\fr{1}{4}F_{\mu\nu}F^{\mu\nu}-\fr{\sgm}{8f}\fr{\eta^{\mu\nu\rho\sigma}}{\sqrt{-g}}F_{\mu\nu}F_{\rho\sigma}\right].
\eeq
Keeping in mind the gauge fixing conditions $A_0=\del_i A_i =0 , ~ h_{ii}=0$, we have the following second order and third order actions
\bea
\label{GFA}S^{(2)}_A &=& \int \d^4 x ~ a^3 \Bigg\{ \fr{1}{2a^2}\dot{A}_i\dot{A}_i -\fr{1}{2a^4}~\del_jA_i~\del_jA_i+\fr{\dot{\sgm}_0}{a^3 f}~\epsilon_{ijk}~\del_j A_k~ A_i\Bigg\},\\
\label{Dc}S_{AA\sgm} &=& \int \d^4 x ~\Bigg\{-\fr{\delta\sgm}{f}~\epsilon_{ijk} ~\dot{A}_i ~\del_j A_k\Bigg\},\\
\label{GI}S_{gAA} &=& \int \d^4 x ~a^3 ~\Bigg\{-\fr{\delta N}{2a^2}\dot{A}_i\dot{A}_i - \fr{\delta N}{4a^4}F_{ij}F_{ij}-\fr{N^{i}}{a^2}\dot{A}_j F_{ij}\Bigg\},\\  
\label{haa}S_{hAA} &=& \int \d^4 x ~a^3 ~\Bigg\{-\fr{1}{2 a^4}h_{ij}\left[a^2 \dot{A}_i\dot{A}_j +\epsilon_{ilm}~\epsilon_{jnp}~\del_lA_m~\del_nA_p\right]\Bigg\}
\eea
Here, through the terms in the action, $S_{gAA}$, more interactions between the scalar sector and the gauge fields arise. First two of these interactions are trivial to write down using \eqref{LS1} in \eqref{GI},
\beq\label{gAA1}
S_{gAA} \supset \int \d^4 x ~a \sum_a \fr{\dot{\phi}_{_{0a}}}{H\Mp^2}\delta \phi_a ~\Bigg\{-\fr{1}{4}\dot{A}_i\dot{A}_i - \fr{1}{8a^2}F_{ij}F_{ij}\Bigg\}.
\eeq  
The last term in \eqref{GI} requires several integration by parts together with use of background equations \eqref{BGX}. This procedure leads to
\beq\label{gAA2}
S_{gAA} \supset \int \d^4 x ~a \sum_a \fr{\dot{\phi}_{_{0a}}}{2H\Mp^2}\delta \phi_a ~\Bigg\{\nabla^{-2}\del_t \del_i\left(\dot{A}_j F_{ij}\right) +H~\nabla^{-2} \del_i\left(\dot{A}_j F_{ij}\right)\Bigg\}.
\eeq
By switching to conformal time $\d \tau =\d t /a$ and using the definitions of Electric and Magnetic fields $B_i = a^{-2}\epsilon_{ijk}\del_jA_k$, $E_i=-a^{-1}\dot{A}_i$, we can re-write the interactions in \eqref{gAA2}, \eqref{gAA1} and \eqref{Dc} in a simpler form
\bea
\label{gi1}S_{gAA} &\supset& -\int \d^3 x~ \d \tau~ a^4 ~ \sum_a \fr{{\phi}'_{_{0a}}}{2\mathcal{H}\Mp^2}~\delta \phi_a ~ \Bigg\{ \fr{\vec{E}^2 +\vec{B}^2}{2}\Bigg\},\\
\label{gi2}S_{gAA} &\supset& -\int \d^3 x~ \d \tau ~ \sum_a \fr{{\phi}'_{_{0a}}}{2\mathcal{H}\Mp^2}~\delta \phi_a ~ \Bigg\{\nabla^{-2}\del_\tau \left(a^4 \vec{\nabla} .(\vec{E}\times\vec{B})\right)\Bigg\},\\
\label{dc}S_{AA\sgm} &=& \int \d^3 x~\d\tau ~a^4 ~\fr{\delta \sgm}{f} \vec{E}.\vec{B}.
\eea
These results agrees with the ones presented in \cite{Barnaby:2011vw}.
\subsection*{Source terms in the equation of motion of $\delta \phi$ and $\delta\sgm$}
In flat gauge there are two gravitationally induced terms that sources the $\delta \phi$ equation. Here we point out a simplification that arise in Fourier space when the two terms are combined. The first interaction can be read easily from the action in \eqref{gi1} which in momentum space reads
\beq\label{J1}
J^{(1)}_{\rm GI}(\vec{k},\tau)=-\fr{\phi_{_0}' a^2}{4\mathcal{H}\Mp^2}\int \fr{\d^3 p}{(2\pi)^{3/2}}\Big[E_i(\vec{k}-\vec{p},\tau)E_i(\vec{p},\tau)+B_i(\vec{k}-\vec{p},\tau)B_i(\vec{p},\tau)\Big].
\eeq
The non-local term in \eqref{gi2} on the other hand requires a bit more work. First note that in configuration space it can be re-written in the following form
\beq\label{J2}
J^{(2)}_{\rm GI} = \fr{\phi_{_0}'}{2\mathcal{H}\Mp^2}\fr{a^2}{\nabla^2}~ \vec{\nabla}.\left(\vec{E}\times(\vec{\nabla}\times \vec{E})+\vec{B}\times(\vec{\nabla}\times\vec{B}) \right),
\eeq
where we have used the equations of motion of electric field together with the Bianchi identity \cite{Anber:2009ua}
\bea
\vec{B}'+ 2\mathcal{H}\vec{B} + \vec{\nabla} \times \vec{E} = 0,\\
\vec{E}'+ 2\mathcal{H}\vec{E} - \vec{\nabla} \times \vec{B} = -\fr{\sgm_{_0}'}{f}\vec{B}.
\eea
On the other hand, in \eqref{J2}, the Fourier transform of the important factors is given by
\beq
\nn \vec{\nabla}.\left(\vec{E}\times(\vec{\nabla}\times \vec{E})\right)(\vec{k})=k_ik_j\int \fr{\d^3 p}{(2\pi)^{3/2}}E_i(\vec{k}-\vec{p})E_j(\vec{p})-k_i \int \fr{\d^3 p}{(2\pi)^{3/2}}(k-p)_i E_j(\vec{k}-\vec{p})E_j(\vec{p}),
\eeq  
where we used the divergenceless condition of electric field in momentum space, $\vec{p}.\vec{E}(\vec{p})=0$ to simplify first term (Note that we suppressed the $\tau$ dependence of the mode functions inside the integrals for the ease of notation). The second integral in the expression above can also be simplified by noticing
\beq
\int \d^3 q ~q_i~ E_j(\vec{k}-\vec{q})E_j(\vec{q}) = \int \d^3 p~(k-p)_i~ E_j(\vec{k}-\vec{p})E_j(\vec{p})
\eeq 
which in turn implies
\beq
 \int \fr{\d^3 p}{(2\pi)^{3/2}}(k-p)_i~ E_j(\vec{k}-\vec{p})E_j(\vec{p}) = \fr{k_i}{2} \int \fr{\d^3 p}{(2\pi)^{3/2}}~ E_j(\vec{k}-\vec{p})E_j(\vec{p}).
\eeq
Taking into account the considerations above, we have
\beq
\vec{\nabla}.\left(\vec{E}\times(\vec{\nabla}\times \vec{E})\right)(\vec{k})= k_ik_j\int \fr{\d^3 p}{(2\pi)^{3/2}}E_i(\vec{k}-\vec{p})E_j(\vec{p})-\fr{k^2}{2} \int \fr{\d^3 p}{(2\pi)^{3/2}}~ E_j(\vec{k}-\vec{p})E_j(\vec{p}).
\eeq
Remembering that ``magnetic'' fields contribute less than ``electric'' fields in our model, we can safely ignore their contribution inside the integrals. Therefore, the Fourier transform of the non-local source term can be written as
\beq
J^{(2)}_{\rm GI}(\vec{k})\simeq \fr{\phi_{_0}'}{2\mathcal{H}\Mp^2}\fr{a^2}{k^2}\Bigg[\fr{k^2}{2} \int \fr{\d^3 p}{(2\pi)^{3/2}}~ E_j(\vec{k}-\vec{p})E_j(\vec{p})- k_ik_j\int \fr{\d^3 p}{(2\pi)^{3/2}}E_i(\vec{k}-\vec{p})E_j(\vec{p})\Bigg]
\eeq
Similarly, ignoring the $\vec{B}$ fields in the expression \eqref{J1}, we combine the two source terms to get a simplified expression
\beq\label{Jt}
J^{(\phi)}_{\rm GI}(\vec{k})\equiv J^{(1)}_{\rm GI}(\vec{k})+J^{(2)}_{\rm GI}(\vec{k})=-\fr{\phi_{_0}'}{2\mathcal{H}\Mp^2}\fr{a^2}{k^2}~k_ik_j\int \fr{\d^3 p}{(2\pi)^{3/2}}E_i(\vec{k}-\vec{p})E_j(\vec{p}).
\eeq
Note that gravity acts democratically on both fluctuations therefore in the e.o.m of the spectator fluctuation $\delta \sgm$, there will be a term similar to \eqref{Jt} that arise from \eqref{J1} and \eqref{J2}. On top of this source term, there is a direct coupling term which can be read from \eqref{dc}. Therefore in total we have the following source term for $\delta \sgm$ in momentum space
\bea
\nn J^{(\sgm)}(\vec{k}) &\simeq & \fr{a^2}{f}\int \fr{\d^3 p}{(2\pi)^{3/2}}E_i(\vec{k}-\vec{p})B_i(\vec{p})-\fr{\sgm_{_0}'}{2\mathcal{H}\Mp^2}\fr{a^2}{k^2}~k_ik_j\int \fr{\d^3 p}{(2\pi)^{3/2}}E_i(\vec{k}-\vec{p})E_j(\vec{p}),\\
&\equiv& J^{(\sgm)}_{\rm D}(\vec{k}) + J^{(\sgm)}_{\rm GI}(\vec{k})
\eea
\section*{Appendix B: Details on the solutions of $A_+$}\label{AppB}

Mode functions of the gauge field satisfy the following equations
\beq
A''_\pm + \left(k^2 \pm 2k \fr{\dot{\sgm}_{_0}}{2Hf\tau}\right) A_\pm = 0, 
\eeq
where we keep the scale factor as $a(\tau) \simeq - (H\tau)^{-1}$, by ignoring sub-leading slow-roll corrections. For $-\infty < \tau < 0 $ and $\dot{\sgm}_{_0} > 0$, positive helicity modes are unstable and the solution that reduces to adiabatic vacuum at early times, \ie $A_+ \to \fr{1}{\sqrt{2k}} e^{-ik\tau}$ as $-k\tau \to \infty$, is given in terms of Coulomb functions
\beq\label{GFS}
A_+(\tau, k) \simeq \fr{1}{\sqrt{2k}}\left[G_0(\xi, -k\tau)+ iF_0(\xi, -k\tau)\right],
\eeq

where $\xi \equiv \dot{\sgm}_{_0}/2Hf$. The approximate equality in \eqref{GFS} arise due to the assumption that the dimensionless measure of field velocity $\xi$ evolves adiabatically, \ie $\dot{\xi}/\xi H \ll 1$, implying
\beq
\fr{\ddot{\sgm}_{_0}}{\dot{\sgm}_{_0} H} -\fr{\dot{H}}{H^2} \ll 1.
\eeq
Further simplifications on the form of the solution \eqref{GFS} arise in the limit where $\xi \gg -k\tau$,
\beq\label{GFS1}
A_+ \simeq \sqrt{\fr{-\tau}{2}}\left[2 e^{\pi\xi}~ \pi^{-1/2} K_1(\sqrt{-8\xi k\tau})+ i  e^{-\pi\xi}~ \pi^{1/2} I_1(\sqrt{-8\xi k\tau})\right],
\eeq
where $I_1$ and $K_1$ are modified Bessel functions of first and second kind. Interesting phenomenology due to gauge field production arise for $\xi \gtrsim \mathcal{O}(1)$, which allows us to further simplify the solution by taking the large argument limit of Bessel function, $-8\xi k\tau \gg 1$,
\beq\label{mfs}
A_+  \simeq \fr{1}{\sqrt{2k}} \left(\fr{-k\tau}{2\xi}\right)^{1/4} e^{\pi\xi -2\sqrt{-2\xi k\tau}} + \fr{i}{\sqrt{2k}} \left(\fr{-k\tau}{2^5\xi}\right)^{1/4} e^{-\pi\xi +2\sqrt{-2\xi k\tau}}.
\eeq
In order to make these approximations to work simultaneously, we require that $\xi > 1/4$. On the other hand, one can further check that these solutions satisfy the condition
\beq\label{mfst}
A'_+ = \sqrt{\fr{2k\xi}{\tau}} A^*_+,
\eeq
corolarly with Wronskian condition $A_+A^{'*}_+ - c.c. =i$. Another important aspect of gauge field production in this model is the fact that growth of $A_+$ modes saturates deep in the IR. This can be seen by taking the limit $-k\tau \to 0$ in \eqref{GFS1} which gives to
\beq
A_+ \rightarrow \fr{e^{\pi\xi}}{2\sqrt{\pi\xi k}}\approx const.
\eeq 
We will see this saturation of the particle production from the perspective of energy density contained in the gauge fields which we now turn in the following section.
\subsection*{Expectation values involving gauge fields}
The expressions we are interested in is the energy density contained in gauge fields $\rho_A$ and the expectation value of the dot product between Electric and Magnetic field $\langle\vec{E}.\vec{B} \rangle$ 
\bea\label{I}
\nn\rho_A \equiv \fr{1}{2}\langle\vec{E}^2+\vec{B^2}\rangle &=& \fr{1}{4\pi^2 a^4}\int \d k\left\{k^2 |A'_+|^2 + k^4|A_+|^2\right\},\\
\langle\vec{E}.\vec{B} \rangle &=& -\fr{1}{4\pi^2 a^4} \int \d k~ k^3 ~\fr{\d}{\d\tau} |A_+|^2 
\eea
For convinience we can write-down both integrands as quantities defined per logarithmic wave-number by using $A_+ = (\sqrt{2k})^{-1} \tilde A(x)$ where $x\equiv -k\tau$, 
\bea
\fr{1}{H^4} \fr{\d \rho_A}{\d \ln k} &=& \fr{x^4}{8\pi^2}\left\{~\left|\fr{\d\tilde A}{\d x}\right|^2 + |\tilde A|^2~\right\},\\
\fr{1}{H^4} \fr{\langle\vec{E}.\vec{B}\rangle}{\d \ln k} &=& \fr{x^4}{8\pi^2}~ \fr{\d}{\d x} |\tilde A|^2
\eea 
Using the approximate solution in \eqref{GFS1} in the $(8\xi)^{-1} \ll -k\tau \ll 2\xi$ regime, we can evaluate the integrals in \eqref{I} analytically. Using the growing Real part of the $A_+$, we can write the energy density as
\beq
\rho_A = \fr{H^4~ e^{2\pi\xi}}{8\pi^2 (2\xi)^{1/2}}\int_0^{2\xi} \d x ~x^{7/2} \left\{\fr{2\xi}{x}+1\right\} e^{-4\sqrt{2\xi x}},
\eeq
where we have set the lower bound of the integral to zero as the integrands quickly vanishes in this limit. Defining a new variable $32\xi x = y^2$, we can re-write the integrals as 
\beq
\rho_A = \fr{H^4}{\xi^3} ~ \fr{e^{2\pi\xi}}{2^{19}\pi^2}\left\{\int_0^{8\xi} \d y ~y^6~ e^{-y}  + \fr{1}{2^6 \xi^2} \int_0^{8\xi} \d y ~ y^{8}~ e^{-y}\right\}.
\eeq
Upper boundary of these integrals can be also sent to infinity $8\xi \to \infty$ in the $\xi \gtrsim \mathcal{O}(1)$ regime as the integrand vanishes quickly for large enough $x$. This gives the result, 
\beq
\rho_A = \fr{H^4}{\xi^3} ~ e^{2\pi\xi}~\fr{\Gamma(7)}{2^{19}\pi^2} \left\{1  + \fr{1}{2^6 \xi^2}\fr{\Gamma(9)}{\Gamma(7)} \right\}.
\eeq
In the $\xi \gtrsim \mathcal{O}(1)$ regime, the first term in the curly brackest dominates which gives the result \eqref{EV} presented in the main text. Following the same steps, one can also obtain the result
\beq
\langle\vec{E}.\vec{B}\rangle = -\fr{H^4}{\xi^4} ~e^{2\pi\xi}~ \fr{\Gamma(8)}{2^{21}\pi^2} \left\{1  - \fr{\Gamma(7)}{\Gamma(8)} \right\}.
\eeq 
\section*{Appendix C: Sourced scalar correlators}\label{AppC}
In this appendix, we derive the sourced scalar and bispectrum in the model \eqref{SF}. A good starting point for this is the source term that appears in the correlators in \eqref{sp} and \eqref{sb}. For this purpose, we first extract the Fourier transforms of $\vec{E}$ and $\vec{B}$ fields from \eqref{EaB} and use the solutions to the mode functions $A_+$ in \eqref{MEAS} to get an explicit expression for the source term in \eqref{s1},
\begin{align}
\nn J_{\vec{k}}(\tau) &= -2 N_{\sgm} \sqrt{\ep\es}\left(\fr{e^{2\pi\xi}}{4~ a(\tau')~f}\right)\int\fr{\d^3p}{(2\pi)^{3/2}} |\vec{k}-\vec{p}|^{1/4}|\vec{p}|^{1/4}~ \left[|\vec{p}|^{1/2}+ |\vec{k}-\vec{p}|^{1/2}\right] \\\nn \\
&~~~~~~~~~~~~~~~~~~~~~~~~~~~~~~~~~~~~~~~~~~~~~~~~~~~\times f(\tau,|\vec{k}-\vec{p}|,|\vec{p}|)~  \hat{\mathcal{O}}_{i,\vec{k}-\vec{p}} ~\hat{\mathcal{O}}_{i,\vec{p}},
\end{align}
where we have symmetrized the integrand with respect to $|\vec{k}-\vec{p}|$ and $|\vec{p}|$ and defined the time dependent function $f$ as 
\beq
f(\tau,|\vec{k}-\vec{p}|,|\vec{p}|) \equiv e^{-2\sqrt{2\xi}~(|\vec{p}|^{1/2}+ |\vec{k}-\vec{p}|^{1/2})\sqrt{-\tau}}.
\eeq
\subsection*{C.1 Power Spectrum}
Using the Wicks theorem for the correlator $\langle\hat{\mathcal{O}}_{i,\vec{k}-\vec{p}} ~\hat{\mathcal{O}}_{i,\vec{p}}~\hat{\mathcal{O}}_{j,\vec{k}'-\vec{p'}} ~\hat{\mathcal{O}}_{j,\vec{p'}}\rangle$, 2-pt. correlator of the source term in \eqref{sp} is given by,
\begin{align}
\nn\langle J_{\vec{k}}(\tau') J_{\vec{k}'}(\tau'')\rangle &= \fr{ N_{\sgm}^2 \ep\es~e^{4\pi\xi}~ \delta(\vec{k}+\vec{k}')}{8~ a(\tau')a(\tau'')~f^2}\int\fr{\d^3p}{(2\pi)^{3}} |\vec{k}-\vec{p}|^{1/2}|\vec{p}|^{1/2}~ \left[|\vec{p}|^{1/2}+ |\vec{k}-\vec{p}|^{1/2}\right]^2\\\nn\\ 
\nn&~~~~~~~~~~~~~~~~~~~~~~~~~~~~~~~~~~\times \left(1+\fr{|\vec{p}|^2-\vec{k}.\vec{p}}{|\vec{k}-\vec{p}||\vec{p}|}\right)^2 f(\tau',|\vec{k}-\vec{p}|,|\vec{p}|)f(\tau'',|\vec{k}-\vec{p}|,|\vec{p}|),
\end{align}
where we have used the following identity for the products of helicity vectors
\beq
|\vec{\epsilon}^{~+}(\vec{k}-\vec{p}).\vec{\epsilon}^{~+}(\vec{p})|^2 =\fr{1}{4}  \left(1+\fr{|\vec{p}|^2-\vec{k}.\vec{p}}{|\vec{k}-\vec{p}||\vec{p}|}\right)^2 .
\eeq

Before we plug the correlators of the sources $\langle J_{\vec{k}} J_{\vec{k}'}\rangle$ in \eqref{sp}, we note that another simplification can be made regarding the Green's functions appearing in this expression: We compute the correlators at late times, $-k\tau \ll 1$ and we observe that the sources associated with $\tau'$ (or $\tau''$) integrals gets most of the contribution from modes with $-k\tau'<\xi^{-1}\ll 1$. Therefore in the $-k\tau<-k\tau'\ll 1$ regime, we can approximate the retarded propagator as
\beq
G^{3/2}_k(\tau,\tau') \simeq -\fr{\tau'^2}{3\tau^2}.
\eeq 
In \eqref{sp}, we therefore have
\begin{align}
\nn \langle \zeta_{\vec{k}}(\tau)\zeta_{\vec{k}'}(\tau)\rangle_{\rm s} &= \fr{H^6}{72~\dot{\phi}_{_0}^2 f^2} N_{\sgm}^2 \ep\es~e^{4\pi\xi}~ \delta(\vec{k}+\vec{k}')\\\nn
&~~\times \int\fr{\d^3p}{(2\pi)^{3}} |\vec{k}-\vec{p}|^{1/2}|\vec{p}|^{1/2}~ \left[|\vec{p}|^{1/2}+ |\vec{k}-\vec{p}|^{1/2}\right]^2 \left(1+\fr{|\vec{p}|^2-\vec{k}.\vec{p}}{|\vec{k}-\vec{p}||\vec{p}|}\right)^2 \\
&~~~~\times\left(\int \d\tau' ~\tau'^3 ~f(\tau',|\vec{k}-\vec{p}|,|\vec{p}|)\right)^2.
\end{align}
The $\tau$ integral above can be integrated easily to give,
\beq\label{Int}
I[z]\equiv \int \d\tau' ~\tau'^3 ~f(\tau',|\vec{k}-\vec{p}|,|\vec{p}|)=-2 \fr{\Gamma(8)}{z^8},
\eeq
where $z \equiv 2\sqrt{2\xi} \left[|\vec{p}|^{1/2}+ |\vec{k}-\vec{p}|^{1/2}\right]$. Noting the relation $\es /f^2 = 2\xi^2 /\Mp^2$, we therefore have 

\begin{align}\label{psa}
\nn \langle \zeta_{\vec{k}}(\tau)\zeta_{\vec{k}'}(\tau)\rangle_{\rm s} &= \fr{H^6}{72~\dot{\phi}_{_0}^2 \Mp^2} \ep N_{\sgm}^2 ~ \fr{\Gamma(8)^2}{2^{21}}~\fr{e^{4\pi\xi}}{\xi^6}~ \delta(\vec{k}+\vec{k}')\\
&~~\times \int\fr{\d^3p}{(2\pi)^{3}} \fr{|\vec{k}-\vec{p}|^{1/2}|\vec{p}|^{1/2}}{\left(|\vec{p}|^{1/2}+ |\vec{k}-\vec{p}|^{1/2}\right)^{14}}~  \left(1+\fr{|\vec{p}|^2-\vec{k}.\vec{p}}{|\vec{k}-\vec{p}||\vec{p}|}\right)^2.
\end{align}
Finally, we define the dimensionless variable $p_* = |\vec{p}|/|\vec{k}|= p/k$ and denote the angle between $\vec{k}$ and $\vec{p}$ by $\theta$ to evaluate the momentum integral numerically. On the other hand, factors of $H^4/\dot{\phi}_{_0}^2$ and $H^2/\Mp^2$ in \eqref{psa} can be expressed in terms of $\Delta^2_{\zeta,\rm v}$ using $\Delta^2_{\zeta,\rm v} = H^4/(4\pi^2 \dot{\phi}_{_0}^2) = H^2/(8\pi^2\ep\Mp^2)$. Putting all the pieces together, we arrive to the following expression
\beq
\langle \zeta_{\vec{k}}(\tau)\zeta_{\vec{k}'}(\tau)\rangle_{\rm s}=\fr{2\pi^2}{k^3}~ 2.9 \times 10^{-4} ~\Big(\ep N_{\sgm}\Delta_{\zeta,\rm v}^2 (k) \Big)^2 ~\fr{e^{4\pi \xi}}{\xi^6}~ \delta(\vec{k}+\vec{k}'). 
\eeq
\subsection*{C.2 Bispectrum}
Following the similar steps in the calculation of power spectrum, we write the correlators of the source that goes in the to the calculation of the bispectrum in \eqref{sb} as
\begin{align}
\nn\langle J_{\vec{k}_1}(\tau_1) J_{\vec{k}_2}(\tau_2)J_{\vec{k}_3}(\tau_3)\rangle &= - \fr{ N_{\sgm}^3 (\ep\es)^{3/2}~e^{6\pi\xi}}{8~ a(\tau_1)a(\tau_2)a(\tau_3)~f^3}\prod_{i=1}^3\int \fr{\d^3 p_i}{(2\pi)^{9/2}} |\vec{k}_i-\vec{p}_i|^{1/4}|\vec{p}_i|^{1/4} \left[|\vec{p}_i|^{1/2}+ |\vec{k}_i-\vec{p}_i|^{1/2}\right]\\\nn
&~~~~~~~~~~~~~~~~~~~~~~~~~~~~~~~~~~~~\times f(\tau_i,|\vec{k}_i-\vec{p}_i|,|\vec{p}_i|)~  \langle\hat{\mathcal{O}}_{l,\vec{k}_i-\vec{p}_i} ~\hat{\mathcal{O}}_{l,\vec{p}_i}\dots\rangle
\end{align} 
Taking the expectation value, \eqref{sb} can be written as 
\begin{align}\label{sba}
\nn \langle \zeta_{\vec{k}_1}(\tau)\zeta_{\vec{k}_2}(\tau)\zeta_{\vec{k}_3}(\tau)\rangle &= -\fr{H^9}{3^3 \dot{\phi}_{_0}^3 f^3} N_{\sgm}^3 (\ep\es)^{3/2}~e^{6\pi\xi}~ \delta(\vec{k}_1+\vec{k}_2+\vec{k}_3)\int \fr{\d^3 p_1}{(2\pi)^{9/2}} \left[ |\vec{k}_1-\vec{p}_1||\vec{p}_1| |\vec{p}_1 +\vec{k}_3| \right]^{1/2}\\\nn
&\times \left[|\vec{p}_1|^{1/2}+ |\vec{k}_1-\vec{p}_1|^{1/2}\right]\left[|\vec{p}_1+\vec{k}_3|^{1/2}+ |\vec{k}_1-\vec{p}_1|^{1/2}\right]
\left[|\vec{p}_1+\vec{k}_3|^{1/2}+ |\vec{p}_1|^{1/2}\right] [\epsilon~ products]\\
&\times \prod_{i=1}^{3} \int \d\tau_i~ \tau_i^3 ~ f(\tau_1,|\vec{k}_1-\vec{p}_1|,|\vec{p}_1|)~ f(\tau_2,|\vec{k}_1-\vec{p}_1|,|\vec{p}_1+\vec{k}_3|)~f(\tau_3,|\vec{p}_1+\vec{k}_3|,|\vec{p}_1|),
\end{align}
where the products of the polarization vectors in the second line of \eqref{sba} can be written as 
\begin{align}
[\epsilon~ products] &\equiv \epsilon^{+ *}_i(\vec{v}_1)\epsilon^{+ }_i(\vec{v}_2) \epsilon^{+ *}_j(\vec{v}_2) \epsilon^{+}_j(\vec{v}_3) \epsilon^{+ *}_k(\vec{v}_3) \epsilon^{+}_k(\vec{v}_1)\\\nn
&=\fr{1}{8}\big[\hat{v}_1.\hat{v}_2+\hat{v}_2.\hat{v}_3+\hat{v}_3.\hat{v}_1 + (\hat{v}_1.\hat{v}_2)^2 +(\hat{v}_2.\hat{v}_3)^2 +(\hat{v}_3.\hat{v}_1)^2\\\nn
&~~~~~~+ (\hat{v}_1.\hat{v}_2)(\hat{v}_2.\hat{v}_3)+(\hat{v}_2.\hat{v}_3)(\hat{v}_3.\hat{v}_1)^2+(\hat{v}_3.\hat{v}_1)(\hat{v}_1.\hat{v}_2)-(\hat{v}_1.\hat{v}_2)(\hat{v}_2.\hat{v}_3)(\hat{v}_3.\hat{v}_1)^2\big].     
\end{align}
Similar to the calculation of the power spectrum, each $\tau$ integral in \eqref{sba} can be calculated analytically. The only difference here is that momentum dependent arguments of the functions $f$ are different so that each integral has to be taken seperately. Doing so, we obtain
\begin{align}\label{sba1}
\nn \langle \zeta_{\vec{k}_1}(\tau)\zeta_{\vec{k}_2}(\tau)\zeta_{\vec{k}_3}(\tau)\rangle &= \fr{H^9}{3^3 \dot{\phi}_{_0}^3 \Mp^3} N_{\sgm}^3 (2\ep)^{3/2}\fr{\Gamma(8)^3}{2^{33}}~\fr{e^{6\pi\xi}}{\xi^9}~ \delta(\vec{k}_1+\vec{k}_2+\vec{k}_3)\int \fr{\d^3 p_1}{(2\pi)^{9/2}}\\\nn 
&\times \fr{ \left[ |\vec{k}_1-\vec{p}_1||\vec{p}_1| |\vec{p}_1 +\vec{k}_3| \right]^{1/2} [\epsilon~ products]}{\left(|\vec{p}_1|^{1/2}+ |\vec{k}_1-\vec{p}_1|^{1/2}\right)^{7}\left(|\vec{p}_1|^{1/2}+ |\vec{p}_1+\vec{k}_3|^{1/2}\right)^{7} \left(|\vec{p}_1+\vec{k}_3|^{1/2}+ |\vec{k}_1-\vec{p}_1|^{1/2}\right)^{7}}
\end{align}
We expect the bispectrum to be maximized in the equilateral configuration, $|\vec{k}_1|=|\vec{k}_2|=|\vec{k}_3|=k$. In this case, to evaluate the momentum integrals, we align $\vec{k}_1$ along the z-axis and define $p_* = |\vec{p}_1|/k$, where $\theta$ is the angle between $\vec{p}_1$ and $\vec{k}_1$, and $\phi$ the angle between the projection of $\vec{p}_1$ and x-direction on the x-y plane,
\begin{align}
\nn\vec{k}_1 = k~(0,0,1),~~ \vec{k}_2 &= k~(-\sqrt{3}/2,0,-1/2),~~ \vec{k}_2 = k~(\sqrt{3}/2,0,-1/2)\\\nn\\
& \vec{p}_1 = kp_* ~\left(\sin\theta\cos\phi,~\sin\theta\sin\phi,~\cos\theta\right).
\end{align}
Given the expressions above \eqref{sba1}, we can evaluate the momentum integral numerically to obtain the bispectrum at the equilateral configuration
\beq
B^{\rm eq}_\zeta (k,k,k) \simeq 10^{-9} (N_{\sgm})^3 \left(\fr{H}{\Mp}\right)^{6} \fr{e^{4\pi\xi}}{\xi^6} \fr{1}{k^6}.
\eeq
\addcontentsline{toc}{section}{References}
\bibliographystyle{utphys}
\bibliography{paper2}

\end{document}